\documentclass[english, citeautoscript, physrev, prb, superscriptaddress, twocolumn]{revtex4-2}
\usepackage[T1]{fontenc}
\usepackage[utf8]{inputenc}
\usepackage{color}
\usepackage{verbatim}
\usepackage{amsmath}
\usepackage{amssymb}
\usepackage{graphicx}
\usepackage{todonotes}

\makeatletter
\usepackage{algorithm,algpseudocode,mathtools}
\usepackage{color}

\makeatother

\usepackage{babel}
\begin{document}
\title{Fast and scalable quantum Monte Carlo simulations of electron-phonon models}
\author{Benjamin Cohen-Stead}
\affiliation{Department of Physics, University of California, Davis, CA 95616,
USA}
\author{Owen Bradley}
\affiliation{Department of Physics, University of California, Davis, CA 95616,
USA}
\author{Cole Miles}
\affiliation{Department of Physics, Cornell University, Ithaca, New York 14853,
USA}
\author{George Batrouni}
\affiliation{\textrm{Université Côte d'Azur, CNRS, Institut de Physique de Nice, (INPHYNI), 06103 Nice, France} }
\affiliation{Centre for Quantum Technologies, National University of Singapore,
2 Science Drive 3, 117542 Singapore}
\affiliation{Department of Physics, National University of Singapore, 2 Science
Drive 3, 117542 Singapore}
\author{Richard Scalettar}
\affiliation{Department of Physics, University of California, Davis, CA 95616,
USA}
\author{Kipton Barros}
\affiliation{Theoretical Division and CNLS, Los Alamos National Laboratory, Los
Alamos, New Mexico 87545, USA}
\begin{abstract}
We introduce methodologies for highly scalable quantum Monte Carlo simulations of electron-phonon models, and report benchmark results for the Holstein model on the square lattice. The determinant quantum Monte Carlo (DQMC) method is a widely used tool for simulating simple electron-phonon models at finite temperatures, but incurs a computational cost that scales cubically with system size. Alternatively, near-linear scaling with system size can be achieved with the hybrid Monte Carlo (HMC) method and an integral representation of the Fermion determinant. Here, we introduce a collection of methodologies that make such simulations even faster. To combat ``stiffness'' arising from the bosonic action, we review how Fourier acceleration can be combined with time-step splitting. To overcome phonon sampling barriers associated with strongly-bound bipolaron formation, we design global Monte Carlo updates that approximately respect particle-hole symmetry. To accelerate the iterative linear solver, we introduce a preconditioner that becomes exact in the adiabatic limit of infinite atomic mass. Finally, we demonstrate how stochastic measurements can be accelerated using fast Fourier transforms. These methods are all complementary and, combined, may produce multiple orders of magnitude speedup, depending on model details.
% The techniques outlined in this paper have been used to generate the results in our accompanying paper \textcolor{red}{[cite Bradley et al., in preparation.]}.
\end{abstract}

% , along with long autocorrelation times,
% have restricted simulations to no more than a few hundred sites
\maketitle
\global\long\def\d{\mathrm{d}}%
\global\long\def\tr{\mathrm{\mathrm{tr}}\,}%
\global\long\def\trep{\mathrm{\mathrm{tr}}_{\textrm{el-ph}}\,}%
\global\long\def\tre{\mathrm{\mathrm{tr}}_{\textrm{el}}\,}%

\section{Introduction}

As a nonperturbative and controlled approach,
quantum Monte Carlo (QMC) methods have been instrumental in advancing
our understanding of interacting solid state systems. In particular,
the broad class of determinant
QMC (DQMC) methods have proven highly effective in helping to characterize
various correlated phases that arise as a result of interactions~\cite{Blankenbecler81}.
Perhaps most notably, DQMC has enabled the study of electron-electron
interactions in the repulsive Hubbard model, where Mott insulator
physics, magnetic order, unconventional superconductivity, and various
additional correlation effects have been observed~\cite{Huang17,Huang18,Li21,Sorella21,Arovas21,Qin21,Scalapino12,Loh05,White89}.
The sign problem, however, has severely limited our ability to simulate
systems without particle-hole or other symmetries, giving rise to an effective
computational cost that scales exponentially with system size and
inverse temperature~\cite{Loh90,Berg12,Chandrasekharan10,Li16,Wu05,Levy21,Tarat22}.

Electron-phonon models, on the other hand, are a family of Hamiltonian
systems that typically evade the sign problem, while still playing
an important role in describing the effect of interactions in solid
state systems. Electron-phonon interactions are essential in explaining
a host of ordered phases in material systems, such as charge density
wave (CDW) order in transition metal dichalcogenides and high
temperature superconductivity in the bismuthates ${\rm Bi}_{1-x}{\rm K}_{x}{\rm BiO}_{3}$~\cite{Jiang21,Li19a,Li20,Sleight15,Xi15,Chen15,Chen16,Wen18,Foyevtsova2015}. Significant
effort has gone towards using DQMC to study Hamiltonian systems with
electron-phonon interactions, in particular the Holstein and Su-Schrieffer-Heeger
(SSH) models~\cite{Bradley21,Cohen-Stead19,Feng20,Li19,Niyaz93,Noack91,Nosarzewski21,Vekic92,Zhang19,Cai21,Feng21,Xing21}.
Although there is no sign problem in such systems, low-temperature DQMC simulations of electron-phonon
models can still be very expensive. Explicit evaluation
of the Fermion determinant results in a computational cost that scales
cubically with system size. Moreover, simulations of both the Holstein
and SSH models suffer from significantly longer autocorrelation times
than comparable DQMC simulations of the repulsive Hubbard model. While DQMC
simulations of the Holstein model have been successfully accelerated
using self-learning Monte Carlo techniques~\cite{Li19c,Xu17}, these gains
are ultimately limited
by continuing to require the evaluation of the Fermion determinant ratio in the Monte Carlo accept/reject step.

In recent years substantial effort has gone towards developing improved
methods for simulating electron-phonon models. Recent work has successfully reduced the computational cost to near linear-scaling in system size. Such scaling can be achieved by avoiding explicit calculation of the Fermion determinant, instead using iterative linear solvers for the sampling and measurement tasks. Applied to simulations of Holstein and SSH models, both Langevin~\cite{Batrouni19,Cohen-Stead20,Goetz21,Karakuzu18} and hybrid Monte Carlo (HMC)~\cite{Beyl18} methods have proven to be highly effective. In this paper, we introduce several general and complementary techniques that can further  reduce the overall costs of simulating electron-phonon models.

Many recent studies of the Holstein and SSH models have used the Langevin method~\cite{Batrouni19,Cohen-Stead20,Goetz21,Karakuzu18}.
The traditional Langevin approach introduces a discretization error associated with the finite time-step used to integrate the stochastic dynamics. Such error can, in principle, be eliminated by introducing an accept/reject step for each proposed Langevin update \cite{Besag94,Rossky78}. An alternative to the Langevin approach is HMC~\cite{Duane87}. Originally developed for lattice gauge theory simulations, the method now finds applications well beyond physics, where HMC also goes by the name {\em Hamiltonian} Monte Carlo~\cite{Neal11}. Interestingly, the Langevin method can be viewed as a special case of HMC, for which the Hamiltonian trajectory length consists of only a single time-step~\cite{Neal11}. Longer trajectories with persistent momentum can be advantageous, however, to reduce autocorrelation times~\cite{Kennedy01}.

As applied to QMC simulations, Langevin and HMC methods offer the promise of near linear-scaling with system size. The general framework is as follows: The aim is to sample a field $x$ according to a probability weight that is proportional to a Fermion determinant $\det M(x)$. Seeking to avoid explicit calculation of this determinant, one instead uses a stochastic approximation scheme, which requires application of the Green function matrix $M^{-1}(x)$ to a vector. The matrix $M(x)$ is highly sparse, and very efficient to apply. Iterative linear solvers, such as conjugate gradient (CG), are effective {\em if} $M(x)$ is well conditioned for typical samples $x$. Good conditioning is not always guaranteed; previous studies of the Hubbard model have found that the condition number can sometimes grow exponentially (e.g., as a function of inverse temperature), making iterative solvers impractical~\cite{Scalettar87,Bai09,Beyl18}. Fortunately, for models of electron-phonon interactions, the condition number of $M(x)$ seems to be reasonably well controlled. Although traditional Langevin and HMC formulations have already been successfully applied to electron-phonon simulation, there are opportunities for substantial improvement, as we shall demonstrate in this paper.

% In lattice gauge theory (LGT) research where near linear scaling with
% system size is required, the community has converged on hybrid Monte
% Carlo (HMC) as the method of choice, also frequently referred to as
% Hamiltonian Monte Carlo~\cite{Duane87,Kennedy01,Neal11}. Like the
% Langevin method, it uses global updates to sample efficiently high
% dimensional spaces with continuous degrees of freedom, but eliminates
% discretization errors, ensuring unbiased statistics. HMC has also
% been used to simulate the Hubbard model, but with mixed results. Technical
% difficulties related to the sign problem and poor numerical conditioning
% have prevented the HMC method from matching the performance of DQMC
% in most cases~\cite{Bai09,Scalettar87}.

% Fortunately, the HMC method seems well suited to simulating electron-phonon
% models, with Ref.~\onlinecite{Beyl18} finding that many of the difficulties
% encountered in simulating the Hubbard model with HMC are absent when
% applied to this new family of Hamiltonian systems. Building on that
% initial effort, we present a refined HMC algorithm for simulating
% electron-phonon models, and apply the method to the widely studied
% Holstein model.

In what follows, we will interweave our new algorithmic developments with benchmarks on a prototypical reference system: the square-lattice Holstein model, which we review in Sec.~\ref{sec:model}. Our core framework for sampling the phonon field is HMC, which we review in Sec.~\ref{sec:hmc}. This application of HMC is fairly sophisticated, involving both Fourier acceleration and time-step splitting to handle the highly disparate time-scales that appear in the bosonic action.

At low temperatures, the sampling of phonons can still be hindered by the formation of tightly-bound bipolarons. To combat this, we employ global Monte Carlo updates as described in Sec.~\ref{sec:special_updates}. For example, by reflecting the entire phonon field ($x \rightarrow -x$) at a particular site, the configuration can ``tunnel through'' a possibly large action barrier. We achieve an improved acceptance rate for these moves by carefully formulating the effective action to respect known particle-hole symmetries of the original Hamiltonian, at least in certain limits. These global updates drastically reduce autocorrelation times, and mitigate ergodicity concerns associated with nodal surfaces (vanishing Fermion determinant)~\cite{Beyl18,Goetz21}, while maintaining excellent scalability of the method.

All components of the simulation can be accelerated by reducing the cost of CG for the linear solves. In Sec.~\ref{sec:Preconditioning} we introduce a preconditioner that significantly reduces the required number of iterations for CG to converge. Specifically, we define the preconditioner $P(x)$ to have the same structure as $M(x)$, but without fluctuations in imaginary time. Application of $P^{-1}(x)$ to a vector can be performed very efficiently through the careful use of the fast Fourier transform (FFT) and Chebyshev polynomial expansion.
% which we review in Appendix~\ref{sec:precond_impl}.

It is important that the computational cost to perform measurements scales like the cost to collect phonon samples, i.e., near-linearly in system size. By Wick's theorem, all electronic measurements can be reduced to products of the single-particle Green function, and the latter can be sampled from the matrix elements $M^{-1}(x)$. It is therefore essential to be able to estimate elements of $M^{-1}(x)$ efficiently. For this we use stochastic techniques that involve applying $M^{-1}(x)$ to random vectors. Section~\ref{sec:measurements} describes how FFTs can be used to achieve near-linear scaling in system size, even when averaging correlation functions over all sites and imaginary-times.

\section{The Holstein model as a benchmark system\label{sec:model}}

\subsection{Model definition}

The methods presented in this paper apply generally to models of electron-phonon interactions, including the SSH and Holstein models. For concreteness, we select the latter for our benchmarks. The Holstein Hamiltonian is~\cite{Holstein59},

\begin{align}
\hat{H} & =\hat{H}_{\mathrm{el}}+\hat{H}_{\mathrm{ph}}+\hat{H}_{\textrm{el-ph}}\label{eq:H_total}\\
\hat{H}_{\mathrm{el}} & =-\sum_{i,j,\sigma} t_{ij}^{\phantom{\dagger}} \hat{c}_{i,\sigma}^{\dagger}
\hat{c}_{j,\sigma}^{\phantom{\dagger}}-\mu\sum_{i,\sigma}\hat{n}_{i,\sigma}\\
\hat{H}_{\mathrm{ph}} & =\frac{m_{\textrm{ph}}\omega_{0}^{2}}{2}\sum_{i}\hat{X}_{i}^{2}+\frac{1}{2m_{\textrm{ph}}}\sum_{i}\hat{P}_{i}^{2}\label{eq:H_ph}\\
\hat{H}_{\textrm{el-ph}} & =\alpha\sum_{i,\sigma}\hat{X}_{i}\left(\hat{n}_{i,\sigma}-\frac{1}{2}\right),\label{eq:H_elph}
\end{align}
with the normalization $\hbar=1$ applied throughout. The first term,
$\hat{H}_{\mathrm{el}}$, models the electron kinetic energy via the
hopping strengths $t_{ij}=t_{ji}$, and controls electron filling through
the chemical potential $\mu$. As usual, $\hat{c}_{i,\sigma}^{\dagger}\big(\hat{c}_{i,\sigma}^{\phantom{\dagger}}\big)$
is the fermionic creation (annihilation) operator for an electron
with spin $\sigma$, and $\hat{n}_{i,\sigma}=\hat{c}_{i,\sigma}^{\dagger}\hat{c}_{i,\sigma}^{\phantom{\dagger}}$
is the electron number operator. The second term, $\hat{H}_{\mathrm{ph}}$,
describes a dispersionless phonon branch with energy $\omega_{0}$ and
mass $m_{\textrm{ph}}$, modeled via the canonical position and momentum
operators $\hat{X}_i$ and $\hat{P}_i$ respectively. Henceforth the atomic mass is normalized
to one, $m_{\textrm{ph}}=1$. The last term, $\hat{H}_{\textrm{el-ph}}$,
introduces an electron-phonon coupling with strength $\alpha$. 

\subsection{Benchmark parameters}

Our methodology applies to models with arbitrary lattice type, hopping matrix,
and electron filling fraction, but we must make some specific choices for our benchmarks. We select the
square lattice Holstein model at half filling ($\mu=0$). We include
only a nearest neighbor electron hopping with amplitude $t_{ij}=1$,
which defines the basic unit of energy. For the square lattice, the
non-interacting bandwidth is then $W=8$. The discretization in imaginary
time, which controls Suzuki-Trotter errors, will be $\Delta_{\tau}=0.1$.
Our benchmarks will vary over the number of lattice sites, $N$, and the inverse temperature, $\beta$. A
useful reference energy scale is the dimensionless electron-phonon
coupling, $\lambda=\alpha^{2}/\big(\omega_{0}^{2}\,W\big)$.
We will consider two coupling strengths, $\lambda=0.25$ or $\lambda=0.60$,
and two phonon frequencies $\omega_{0}=0.1$ and $\omega_{0}=1$. 

For these Holstein systems, the stable phase at low temperatures and half-filling is
charge-density-wave (CDW) order; electrons form a checkerboard pattern,
spontaneously breaking the $\mathbb{Z}_{2}$ symmetry between sublattices.
In the case of $\omega_{0}=1.0$ and $\lambda=0.25$,
the CDW transition temperature is $\beta_{{\rm cdw}}\approx6$
\cite{Batrouni19,Costa18}.
To detect this phase, we measure the $\left(\pi,\pi\right)$ charge structure
factor
\begin{align}
S_{{\rm cdw}} & =\sum_{\mathbf{r}}\left(-1\right)^{r_{x}+r_{y}}C(\mathbf{r}),\label{eq:Scdw}
\end{align}
where
\begin{equation}
C(\mathbf{r})=\frac{1}{N}\sum_{\mathbf{r}'}\left\langle \hat{n}_{\mathbf{r}'+\mathbf{r}}\hat{n}_{\mathbf{r}'}\right\rangle ,
\end{equation}
is the real-space density-density correlations in $\hat{n}_{\mathbf{r}}=\hat{n}_{\mathbf{r},\uparrow}+\hat{n}_{\mathbf{r},\downarrow}$.
Here we are using integers $\mathbf{r}=(r_{x},r_{y})$ to index sites
on the square lattice, assuming periodic boundary conditions. Superconducting
order, on the other hand, can be detected using the pair susceptibility
\begin{equation}
P_{s}=\frac{1}{N}\int_0^\beta \left\langle \hat{\Delta}\left(\tau\right)\hat{\Delta}^{\dagger}\left(0\right)\right\rangle d\tau,
\end{equation}
where $\hat{\Delta}\left(\tau\right)=\sum_{\mathbf{r}}\hat{c}_{\mathbf{r},\downarrow}\left(\tau\right)\hat{c}_{\mathbf{r},\uparrow}\left(\tau\right)$.

All results reported in this paper use HMC trajectories comprised
of $N_{t}=100$ time-steps (Sec.~\ref{sec:hmc}). Except where noted,
we will use Fourier acceleration with mass regularization $m_{\mathrm{reg}}=\omega_{0}$
(Sec.~\ref{subsec:mass_matrix}), and time-step splitting with $n_{t}=10$
(Sec.~\ref{subsec:Time-step-splitting}). We will use a varying number
of thermalization and simulation HMC trial updates, denoted $N_{{\rm therm}}$
and $N_{{\rm sim}}$, respectively, with measurements taken after
each simulation update. 

\subsection{Path integral representation}

To measure thermodynamic properties, one can formulate a path integral
representation of the partition function. A full derivation is given
in Appendix~\ref{sec:path_integral}, with the result
\begin{align}
\mathcal{Z} & =\trep e^{-\beta\hat{H}}\nonumber \\
 & \approx\int\mathcal{D}x\,e^{-\left(S_{{\rm B}}-\Delta_{\tau}\alpha\sum_{i,\tau}x_{i,\tau}\right)}\left(\det M\right)^{2}.\label{eq:Z_path_orig}
\end{align}
Here the inverse temperature $\beta$ has been discretized into $L_{\tau}$
intervals of imaginary time, with $\Delta_{\tau}=\beta/L_{\tau}$. The
integral goes over all sites $i$ and imaginary times $\tau$ in the
real phonon field $x_{i,\tau}$. The ``bosonic action''
\begin{equation}
S_{\mathrm{B}}=\frac{\Delta_{\tau}}{2}\sum_{i=1}^{N}\sum_{\tau=0}^{L_{\tau}-1}\left[\omega_{0}^{2}x_{i,\tau}^{2}+\left(\frac{x_{i,\tau+1}-x_{i,\tau}}{\Delta_{\tau}}\right)^{2}\right],\label{eq:S_B}
\end{equation}
describes dispersionless phonon modes, but can be readily generalized
to include anharmonic terms and phonon dispersion~\cite{Costa18,Paleari21}.
The ``Fermion determinant'' involves the $NL_{\tau}\times NL_{\tau}$
matrix,

\begin{equation}
M=\left(\begin{array}{ccccc}
I &  &  &  & B_{0}\\
-B_{1} & I\\
 & -B_{2} & \ddots\\
 &  & \ddots & \ddots\\
 &  &  & -B_{L_{\tau}-1} & I
\end{array}\right),\label{eq:M_def}
\end{equation}
comprised of $N\times N$ blocks. The off-diagonal blocks are
\begin{align}
B_{\tau} & =e^{-\Delta_{\tau}V_{\tau}}e^{-\Delta_{\tau}K},\label{eq:B_def}
\end{align}
where the matrices
\begin{equation}
\left(V_{\tau}\right)_{ij}=\delta_{ij}\left(\alpha x_{i,\tau}-\mu\right),\qquad K_{ij}=-t_{ij},\label{eq:V_K_def}
\end{equation}
describe the electron-phonon coupling and the electron hopping, respectively.
In this real-space basis, $e^{-\Delta_{\tau}V_{\tau}}$ is exactly
diagonal, whereas $e^{-\Delta_{\tau}K}=I-\Delta_{\tau}K+\dots$ is highly sparse up to corrections of order $\Delta_{\tau}^2$. Note that one could alternatively formulate~\cite{Blankenbecler81}
\begin{equation}
\det M=\det(I+B_{L_{\tau}-1}\dots B_{1}B_{0}),\label{eq:detM_alt}
\end{equation}
but we do not pursue that approach here.

An innovation in this work is to rewrite the partition function as
\begin{equation}
\mathcal{Z}\approx\int\mathcal{D}x\,e^{-S_{{\rm B}}}\left[\det\left(M\Lambda\right)\right]^{2},\label{eq:Z_path}
\end{equation}
where $\Lambda(x)$ is any matrix that satisfies
\begin{equation}
\det\Lambda^{2}=e^{\Delta_{\tau}\alpha\sum_{i,\tau}x_{i,\tau}}.\label{eq:Lambda_constraint}
\end{equation}
Although Eqs.~(\ref{eq:Z_path_orig}) and~(\ref{eq:Z_path}) are
mathematically equivalent, this reformulation will have important
consequences in Secs.~\ref{subsec:aux} and~\ref{sec:special_updates}.
The factor $\exp(\Delta_{\tau}\alpha\sum_{i,\tau}x_{i,\tau})$ originates
from our choice to include the $-\alpha\sum_{i}\hat{X}_{i}/2$ term
in Eq.~(\ref{eq:H_elph}), which effectively selects $\hat{X}_{i} = 0$ as the reflection point for particle-hole symmetry.

There are many possible choices for $\Lambda$. We select
\begin{equation}
\Lambda_{\left(i,\tau\right),\left(i',\tau'\right)}=\delta_{i,i'}\delta_{\tau+1,\tau'}\left(2\delta_{\tau',0}-1\right)e^{+\frac{\Delta_{\tau}\alpha}{2}x_{i,\tau'}},\label{eq:Lambda_def}
\end{equation}
with inverse
\begin{equation}
\Lambda_{\left(i,\tau\right),\left(i',\tau'\right)}^{-1}=\delta_{i,i'}\delta_{\tau,\tau'+1}\left(2\delta_{\tau,0}-1\right)e^{-\frac{\Delta_{\tau}\alpha}{2}x_{i,\tau}},\label{eq:Lambda_inv}
\end{equation}
where the index $\tau=0,1,\dots L_{\tau}-1$ is understood to be periodic
in $L_{\tau}$.
% Consequence of this definition for $\Lambda$ will be discussed in Sec.~\ref{sec:special_updates}.

To collect equilibrium statistics, one samples the phonon field $x_{i,\tau}$,
taking the positive-definite integrand in Eq.~(\ref{eq:Z_path})
to be the probability weight. Sampling $x_{i,\tau}$ is typically
the dominant cost of a QMC code. A traditional DQMC code involves periodic evaluation of the matrix determinant of Eq.~(\ref{eq:detM_alt}), at a cost of $\mathcal{O}(N^{3})$ computational operations. In a careful DQMC implementation, this determinant may be calculated relatively infrequently, typically once per ``full sweep'' of Monte Carlo updates to each of the auxiliary field components, $x_{i,\tau}$~\cite{Gubernatis16}. 
As we will next discuss in Sec.~\ref{subsec:aux}, the cost to sample
the phonon field can still be significantly reduced, from cubic to approximately linear scaling with
system size $N$.

% Samples of $x_{i,\tau}$ provide estimates of the time-dependent Green function via the matrix elements $G_{(i,\tau)(j,\tau')}=M^{-1}_{(i,\tau)(j,\tau')}$. The efficient calculation of observables is the
% subject of Sec.~\ref{sec:measurements}.

Note that a similar path integral formulation can be derived for the
SSH model. There, however, the phonon position operators $\hat{X}_i$ modulate the electron hopping term, such that the matrices $K_{\tau}$ gain a dependence on the phonon fields $x_{i,\tau}$~\cite{Li20,Su79,Cai21,Feng21,Goetz21,Xing21,Beyl18}.

\subsection{Sampling the phonon field at approximately linear scaling cost\label{subsec:aux}}

Given a non-singular matrix $A$ of dimension $D$, its determinant
can be formulated as an integral,
\begin{equation}
\left|\det A\right|=\left(2\pi\right)^{-D/2}\int\mathcal{D}\Phi\,e^{-\frac{1}{2}\Phi^{T}\left(A^{T}A\right)^{-1}\Phi},\label{eq:Phi_identity}
\end{equation}
where each component of the vector $\Phi$ is understood to be integrated
over the entire real line.

We twice apply Eq.~(\ref{eq:Phi_identity}) to Eq.~(\ref{eq:Z_path}), introducing an integral for each of the two Fermion determinants. Taking
\begin{equation}
A(x)=M(x)\Lambda(x),
\end{equation}
the partition
function becomes
\begin{equation}
\mathcal{Z}\approx\left(2\pi\right)^{NL_{\tau}}\int\mathcal{D}\Phi_{\uparrow}\mathcal{D}\Phi_{\downarrow}\mathcal{D}x\,e^{-S(x,\Phi_{\sigma})}.
\end{equation}
In place of the matrix determinants, there is now a ``fermionic''
contribution to the action,
\begin{equation}
S(x,\Phi_{\sigma})=S_{\mathrm{B}}(x)+S_{\mathrm{F}}(x,\Phi_{\sigma}),\label{eq:S_decompose}
\end{equation}
with
\begin{align}
S_{\mathrm{F}}\left(x,\Phi_{\sigma}\right) & =\frac{1}{2}\sum_{\sigma}\Phi_{\sigma}^{T}\left(A^{T}A\right)^{-1}\Phi_{\sigma}\nonumber \\
 & =\frac{1}{2}\sum_{\sigma}\left|A^{-T}\Phi_{\sigma}\right|^{2}.\label{eq:S_F}
\end{align}
Now we must sample the two auxiliary fields $\Phi_{\{\uparrow,\downarrow\}}$
in addition to the phonon field $x$, according to the joint distribution
$P(x,\Phi_{\sigma})\propto\exp(-S)$. With the Gibbs sampling method,
one alternately updates $x$ and $\Phi_{\sigma}$ according to the
conditional distributions $P\left(x\vert\Phi_{\sigma}\right)$ and
$P\left(\Phi_{\sigma}\vert x\right)$ respectively. 

Holding $x$ fixed, observe that 
\begin{equation}
P(\Phi_{\sigma}|x)\propto e^{-S_{F}}=e^{-\frac{1}{2}\sum_{\sigma}\left|R_{\sigma}\right|^{2}},
\end{equation}
where the vector $R_{\sigma} = A^{-T} \Phi_\sigma$ is found to be Gaussian distributed. Therefore, to sample $\Phi_{\sigma}$ at fixed $x$, one may first
sample Gaussian $R_{\sigma}$, and then assign
\begin{equation}
\Phi_{\sigma} = A^{T}(x) R_{\sigma}.\label{eq:PhiSample-1}
\end{equation}
Because $\Phi_\sigma$ is randomly sampled, it is convenient to treat it as an arbitrary, fixed vector. Alternatively, we can view $\Phi_{\sigma}(x,R_\sigma)$ as a deterministic function of $x$ provided that the random sample $R_\sigma$ is also supplied.

Sampling $x$ at fixed $\Phi_{\sigma}$ is the primary numerical challenge. In the
Metropolis Monte Carlo approach, one proposes an update $x\rightarrow x'$
and accepts it with probability,
\begin{equation}
P(x\rightarrow x')=\min\left(1,e^{-\Delta S}\right),\label{eq:metropolis}
\end{equation}
where
\begin{equation}
\Delta S=S(x',\Phi_{\sigma})-S(x,\Phi_{\sigma}).
\end{equation}
Sophisticated methods for proposing updates include HMC (Sec.~\ref{sec:hmc})
and reflection/swap updates (Sec.~\ref{sec:special_updates}).

Calculating the acceptance probability requires evaluating the change
in action,
\begin{equation}
\Delta S=\Delta S_{\mathrm{B}}+\Delta S_{\mathrm{F}}.\label{eq:Delta_S_F}
\end{equation}
The bosonic part can be readily calculated from Eq.~(\ref{eq:S_B}).
The fermionic part is given by Eq.~(\ref{eq:S_F}),
\begin{equation}
\Delta S_{\mathrm{F}} =S_{\mathrm{F}}(x',\Phi_{\sigma})-S_{\mathrm{F}}(x,\Phi_{\sigma}).
\end{equation}
The recipe for sampling the auxiliary field $\Phi_\sigma = \Phi_\sigma(x,R_\sigma)$ is given by Eq.~(\ref{eq:PhiSample-1}), and involves the {\em initial} phonon configuration $x$.
% Specifically, $\Phi_\sigma = A^T(x) R_\sigma$, where $R_\sigma$ is a known vector, sampled from the Gaussian distribution.
Substituting into Eq.~(\ref{eq:S_F}) yields $S_{\mathrm{F}}(x,\Phi_{\sigma}) = \frac{1}{2}\sum_{\sigma}\left|R_{\sigma}\right|^{2}$.

It remains nontrivial to calculate
\begin{equation}
S_{\mathrm{F}}(x',\Phi_{\sigma})=\frac{1}{2}\sum_{\sigma}\Phi_{\sigma}^{T}\Psi_{\sigma},\label{eq:Sf_Psi}
\end{equation}
where
\begin{align}
\Psi_{\sigma} & =\left(A^{T}A\right)^{-1}\Phi_{\sigma}\nonumber \\
 & =\Lambda^{-1}\left(M^{T}M\right)^{-1}\Lambda^{-T}\Phi_{\sigma},\label{eq:psi_def}
\end{align}
and the matrices $M$ and $\Lambda$ are understood to be evaluated
at the new phonon field, $x'$. The vector $b=\Lambda^{-T}\Phi_{\sigma}$, for each $\sigma$,
can be readily calculated using Eq.~(\ref{eq:Lambda_inv}).

To solve iteratively for the vector
\begin{equation}
v=\left(M^{T}M\right)^{-1}b,\label{eq:CG}
\end{equation}
one can use the conjugate gradient (CG) method~\cite{Saad03}.
After $n$ iterations, CG optimally approximates $v_{n}\approx v$
from within the $n$th Krylov space, i.e.~the vector space spanned
by basis vectors $\big(M^{T}M\big)^{j}b$ for $j=0,1,\dots n$.
Given the solution $v$, the action $S_{\rm F}$ in Eq.~(\ref{eq:Sf_Psi}) can be evaluated by noting that $\Phi_{\sigma}^{T}\Psi_{\sigma}=b^{T}v$.

CG requires repeated multiplication by $M^{T}M$. Applying $M$ and
$M^{T}$ to a vector is very efficient due to the block sparsity
structure in Eq.~(\ref{eq:M_def}). The off-diagonal blocks $B_{\tau}$
inside $M$ involve the exponential of the tight-binding hopping matrix
$K$. To apply efficiently $e^{-\Delta_{\tau}K}$ to a vector, one
may approximately factorize this exponential as a chain of sparse
operators using the minimal split checkerboard method~\cite{Lee13},
which remains valid up to errors of order $O\big(\Delta_{\tau}^{2}\big)$
\cite{Beyl20}. This allows us to apply $B_{\tau}$ to a vector
of like dimension at a cost that scales linearly with system size
$N$.

The rate of CG convergence is determined by the condition number of
$M^{T}M,$ i.e., the ratio of largest to smallest eigenvalues (as
a function of the fluctuating phonon field). In previous QMC studies
on the Hubbard model, the analogous condition number was found to
increase rapidly with inverse temperature and system size at moderate coupling~\cite{Bai09}.
Fortunately, for electron-phonon models at moderate parameter values ($\lambda \lesssim 1$ and $\omega_{0}\lesssim t$)
the condition number is observed to increase only very slowly
with $\beta$ and $N$~\cite{Beyl18}. We observe that larger phonon frequency $(\omega_0 \gtrsim t)$ coincides with larger condition number and slower CG convergence.
This will be reflected by the benchmarks in this paper, for which CG typically converges in hundreds of iterations or fewer.
Furthermore, the required number of CG iterations can be significantly
reduced by using a carefully designed preconditioning matrix, as we
will describe in Sec.~\ref{sec:Preconditioning}.

\section{HMC sampling of the phonon field\label{sec:hmc}}

\subsection{Review of HMC}

Hybrid Monte Carlo (HMC) was originally developed in the lattice gauge
theory community~\cite{Duane87}, and has since proven broadly
useful for statistical sampling of continuous variables~\cite{Neal11}.
In particular, it is a powerful method for sampling the phonon field
$x$ in electron-phonon models~\cite{Beyl18,Scalettar88}.

In HMC a fictitious momentum $p_{i,\tau}$ is introduced that is dynamically
conjugate to $x_{i,\tau}$. Specifically, a Hamiltonian
\begin{equation}
H(x,p)=S\left(x\right)+\frac{p^{T}\mathcal{M}^{-1}p}{2},
\end{equation}
is defined that can be interpreted as the sum of ``potential''
and ``kinetic'' energies. The dynamical mass $\mathcal{M}$ can
be any positive-definite matrix, independent of $x$ and $p$. Recall
that the action $S(x)$ is implicitly dependent on the auxiliary field
$\Phi_{\sigma}$; we omit this dependence because $\Phi_{\sigma}$
is treated as fixed for purposes of sampling $x$.

The corresponding Hamiltonian equations of motion are
\begin{align}
\dot{p}=-\frac{\partial H}{\partial x}= & -\frac{\partial S}{\partial x}\label{eq:dyn1}\\
\dot{x}=\frac{\partial H}{\partial p}= & \mathcal{M}^{-1}p.\label{eq:dyn2}
\end{align}
The dynamics is time-reversible, energy conserving, and symplectic
(phase space volume conserving). These properties make it well suited
for proposing updates to the phonon field. We use a variant of HMC
consisting of the following three steps:

Step (1) of HMC samples $p$ from the equilibrium Boltzmann distribution,
proportional to $\exp(-p^{T}\mathcal{M}^{-1}p/2)$. This is achieved
by sampling components $R_{i,\tau}$ from a standard Gaussian distribution,
and then setting 
\begin{equation}
p=\sqrt{\mathcal{M}}R.\label{eq:sample_p}
\end{equation}

Step (2) of HMC integrates the Hamiltonian dynamics for $N_{t}$
integration time-steps. We use the leapfrog method,
\begin{align}
p_{t+1/2} & =p_{t}-\frac{\Delta t}{2}\frac{\partial S}{\partial x_{t}}\label{eq:leapfrog1}\\
x_{t+1} & =x_{t}+\Delta t \mathcal{M}^{-1} p_{t+1/2}\\
p_{t+1} & =p_{t+1/2}-\frac{\Delta t}{2}\frac{\partial S}{\partial x_{t+1}},\label{eq:leapfrog3}
\end{align}
where $\Delta t$ denotes the integration step size. Note that when
performing leapfrog integration steps sequentially, only a single
evaluation of $\partial S/\partial x$ must be performed per time-step.
This is because the final half-step momentum update $p_{t+1/2}\rightarrow p_{t+1}$
can be merged with the initial one from the next time-step, $p_{t'}\rightarrow p_{t'+1/2}$,
where $t'=t+1$. The leapfrog integration scheme is exactly time-reversible
and symplectic. One integration step is accurate to $\mathcal{O}(\Delta t^{3})$
and, in the absence of numerical instability, total energy is conserved to order
$\mathcal{O}(\Delta t^{2})$ for arbitrarily long trajectories~\cite{Clark07,Kennedy13,Leimkuhler04}. Any symplectic integration scheme could be used in place of leapfrog; the second-order Omelyan integrator is an especially promising alternative~\cite{Sexton92,Takaishi06}.

For this paper we will fix $N_{t}=100$. Future work would likely benefit from randomizing the length of each HMC trajectory; doing so has been observed to reduce decorrelation times and would mitigate certain ergodicity concerns~\cite{Mackenze89}. 
For certain types of sampling problems, e.g. sampling in the vicinity of a critical point, one should also consider the use of much longer HMC trajectories~\cite{Kennedy01}.

Step (3) of HMC is to accept (or reject) the dynamically evolved configuration
$x'$ according to the Metropolis probability, Eq.~(\ref{eq:metropolis}).
HMC exactly satisfies detailed balance, and the proof depends
crucially on the leapfrog integrator being time-reversible and symplectic~\cite{Duane87,Neal11}.
An acceptance rate of order one can be maintained by taking the timestep 
to scale only very weakly with system size ($\Delta t\sim N^{-1/4}$)~\cite{Creutz88}. Higher order symplectic integrators are also
possible, and come even closer to allowing constant $\Delta t$, independent of system size~\cite{Clark07}. Extensive discussion about selecting a good $\Delta t$ value is presented in Ref.~\onlinecite{Neal11}, including evidence that a 65\% acceptance rate may be close to optimal in many situations. In practice, however, we typically choose a more conservative $\Delta t$ such that the acceptance rate is much closer to one. This minimizes the danger of hitting numerical instabilities, which can lead to significant slow-downs, and can be difficult to anticipate over a wide space of physical parameters.

Numerical integration requires evaluation of the fictitious force
$-\partial S/\partial x$ at each time-step.
Specifically, one must calculate
\begin{equation}
\frac{\partial S}{\partial x_{i,\tau}}=\frac{\partial S_{{\rm B}}}{\partial x_{i,\tau}}+\frac{\partial S_{{\rm F}}}{\partial x_{i,\tau}}.
\end{equation}
The bosonic part is
\begin{equation}
\frac{\partial S_{{\rm B}}}{\partial x_{i,\tau}}=\Delta_{\tau}\left(\omega_{0}^{2}x_{i,\tau}-\frac{x_{i,\tau+1}-2x_{i,\tau}+x_{i,\tau-1}}{\Delta{}_{\tau}^{2}}\right).\label{eq:dSB_dx}
\end{equation}
For the fermionic part, we must calculate
\begin{equation}
\frac{\partial S_{{\rm F}}}{\partial x_{i,\tau}}=\frac{1}{2}\sum_{\sigma}\Phi_{\sigma}^{T}\frac{\partial(A^{T}A)^{-1}}{\partial x}\Phi_{\sigma},
\end{equation}
where $\Phi_{\sigma}$ is fixed throughout the dynamical trajectory. Using the general matrix identity
$\mathrm{d}C^{-1}=-C^{-1}(\mathrm{d}C)C^{-1}$, we find
\begin{equation}
\frac{\partial S_{{\rm F}}}{\partial x_{i,\tau}}=-\sum_{\sigma}\Psi_{\sigma}^{T}A^{T}\frac{\partial A}{\partial x_{i,\tau}}\Psi_{\sigma},\label{eq:dSF_dx}
\end{equation}
where $\Psi_{\sigma}=\left(A^{T}A\right)^{-1}\Phi_{\sigma}$ evolves as a function of $x$ over the dynamical trajectory, with $\Phi_{\sigma}$ held fixed. As with the calculation
of $\Delta S_{F}$ in Eq.~(\ref{eq:Delta_S_F}), the numerically
expensive task is to calculate $\Psi_{\sigma}=\left(A^{T}A\right)^{-1}\Phi_{\sigma}$,
for which we use the CG algorithm.

Given $\Psi_{\sigma}$, we must also apply the highly sparse matrix
\begin{equation}
\frac{\partial A_{x}}{\partial x_{i,\tau}}=\frac{\partial M}{\partial x_{i,\tau}}\Lambda+M\frac{\partial\Lambda}{\partial x_{i,\tau}},
\end{equation}
for each index $(i,\tau)$ of the phonon field. Differentiating $\Lambda$
in Eq.~(\ref{eq:Lambda_def}) is straightforward. The derivative
of $M$ in Eq.~(\ref{eq:M_def}) with respect to $x_{i,\tau}$ involves
only a single nonzero $N\times N$ block matrix. In the Holstein model,
we use
\begin{equation}
\frac{\partial B_{\tau'}}{\partial x_{i,\tau}}=\delta_{\tau,\tau'}\left(\frac{\partial}{\partial x_{i,\tau}}e^{-\Delta_\tau V_{\tau}}\right)e^{-\Delta_{\tau}K},
\end{equation}
where $V_{\tau}$ is diagonal, so that its exponential is easy to
construct and differentiate.

The situation is a bit more complicated for the SSH model, where
the $x_{i,\tau}$-dependence appears inside the
hopping matrix $K_{\tau}$, which is \emph{not} diagonal. In this
case, we may exploit the checkerboard factorization~\cite{Lee13}
of $e^{-\Delta_{\tau}K_{\tau}}$, and use the product rule to differentiate
each of the sparse matrix factors one-by-one. If implemented carefully,
the cost to evaluate all $N\, L_{\tau}$ forces $-\partial S/\partial x_{i,\tau}$
remains of the same order as the cost to evaluate the scalar $S$.
That this is generically possible follows from the concepts of reverse-mode
automatic differentiation~\cite{Griewank89}.

\subsection{Resolving disparate time-scales in the bosonic action}

One of the challenges encountered when simulating electron-phonon
models is that the bosonic action gives rise to a large disparity
of time-scales in the Hamiltonian dynamics. Here we will present two
established approaches for unifying these dynamical time scales.

The bosonic part of the Hamiltonian dynamics decouples in the Fourier
basis. To see this, we will employ the discrete Fourier transform
in imaginary time,
\begin{align}
\hat{f}_{\omega} & =\frac{1}{\sqrt{L_{\tau}}}\sum_{\tau=0}^{L_{\tau}-1}e^{-\frac{2\pi\mathrm{i}}{L_{\tau}}\omega\tau}f_{\tau}.\label{eq:F_forward}
\end{align}
where the integer index $\omega$ is effectively periodic mod $L_{\tau}$.
The Fourier transform may be represented by an $L_{\tau}\times L_{\tau}$
unitary matrix,
\begin{equation}
\mathcal{F}_{\omega,\tau}=\frac{1}{\sqrt{L_{\tau}}}e^{-\frac{2\pi\mathrm{i}}{L_{\tau}}\omega\tau},\label{eq:F_def}
\end{equation}
such that $\hat{f}=\mathcal{F}f$.

Consider the bosonic force defined in Eq.~(\ref{eq:dSB_dx}), 
\begin{equation}
f_{i,\tau}=-\partial S_{\mathrm{B}}/\partial x_{i,\tau}.
\end{equation}
Its Fourier transform is
\begin{equation}
\hat{f}_{i,\omega}=-\tilde{Q}_{\omega,\omega}\hat{x}_{i,\omega},\label{eq:f_hat_force}
\end{equation}
where $\hat{x}=\mathcal{F}x$ and
\begin{equation}
\tilde{Q}_{\omega,\omega}=\Delta_{\tau}\left[\omega_{0}^{2}+\frac{4}{\Delta_{\tau}^{2}} \sin^2 \left( \frac{2\pi\omega}{L_{\tau}}\right)\right].\label{eq:Qww_def}
\end{equation}
We may interpret $\tilde{Q}_{\omega,\omega}$ as the elements of a
diagonal matrix $\tilde{Q}$ in the Fourier basis. In the original
basis,
\begin{equation}
\partial S_{\mathrm{B}}/\partial x=Qx,\label{eq:bosonic_force}
\end{equation}
where $Q=\mathcal{F}^{-1}\tilde{Q}\mathcal{F}$.

The diagonal matrix element $\tilde{Q}_{\omega,\omega}$ gives the force
acting on the Fourier mode $\hat{x}_{\omega}$. The extreme cases
are $\omega=\pm L_{\tau}/2$ and $\omega=0$, for which $\tilde{Q}_{\omega,\omega}/\Delta_{\tau}$
takes the values $\omega_{0}^{2}+4/\Delta_{\tau}^{2}$ and $\omega_{0}^{2}$
respectively. The ratio of force magnitudes for the fastest and slowest
dynamical modes is then
\begin{equation}
1+\frac{4}{\omega_{0}^{2}\Delta_{\tau}^{2}}\gg1,
\end{equation}
which diverges in the continuum limit, $\Delta_{\tau}\rightarrow0$.
Typically $\Delta_{\tau}$ is of order $0.1$, and the physically
relevant phonon frequencies are order $\omega_{0}\sim0.1$.

Numerical integration of the Hamiltonian dynamics will be limited
to small time-steps to resolve the dynamics of the fast modes, $\omega\sim\pm L_{\tau}/2$.
Unfortunately, this means that a very large number of time steps $N_{t}\propto\mathcal{O}\left(4/\omega_{0}^{2}\Delta_{\tau}^{2}\right)$
is required to reach the dynamical time-scale in which the slow modes,
$\omega\sim0$, can meaningfully evolve.

\subsubsection{Dynamical mass matrix\label{subsec:mass_matrix}}

Here we describe the method of Fourier acceleration, by which a careful
selection of the dynamical mass matrix $\mathcal{M}$ can counteract
the widely varying bosonic force scales appearing in Eq.~(\ref{eq:Qww_def})
\cite{Batrouni19,Batrouni85}.

The Hamiltonian dynamics of Eqs.~(\ref{eq:dyn1}) and~(\ref{eq:dyn2})
may be written $\ddot{x}=-\mathcal{M}^{-1}\partial S/\partial x$.
The characteristic scaling for fermionic forces is 
\begin{equation}
\partial S_{\mathrm{F}}/\partial x\sim\Delta_{\tau}.\label{eq:SF_scaling}
\end{equation}
This is expected because $\Delta_{\tau}$ enters into $S_{\mathrm{F}}$
only through the scaled phonon field, $y_{i,\tau}\equiv\Delta_{\tau}x_{i,\tau}$.
The chain rule $\partial S_{\mathrm{F}}/\partial x=\left(\partial S_{\mathrm{F}}/\partial y\right)\left(\partial y/\partial x\right)$
then suggests linear scaling in $\Delta_{\tau}$.

Per Eq.~(\ref{eq:Qww_def}), the bosonic forces also typically scale
like $\Delta_{\tau}$ when $\omega$ is small. However, for the large
Fourier modes $\omega\sim\pm L_{\tau}/2$, we find instead
\begin{equation}
\partial S_{\mathrm{B}}/\partial x\sim\Delta_{\tau}^{-1},\label{eq:SB_scaling}
\end{equation}
which will typically dominate other contributions to the total force.
One may therefore consider the idealized limit of a purely bosonic
action, $S(x)=S_{\mathrm{B}}(x)$, which is approximately valid for
the large $\omega$ modes. Using Eq.~(\ref{eq:bosonic_force}), the
dynamics for purely bosonic forces is
\begin{equation}
\ddot{x}=-\mathcal{M}^{-1}Qx\quad\quad(S=S_{\mathrm{B}}),\label{eq:dyn_bos}
\end{equation}
If we were to select $\mathcal{M}=Q/\omega_{0}^{2}$, then the dynamics
would become $\ddot{\hat{x}}=-\omega_{0}^{2}\hat{x}$, which describes
a system of non-interacting harmonic oscillators, all sharing the
\emph{same} period, $2\pi/\omega_{0}$. This would be the ideal choice
of $\mathcal{M}$ if the assumption $S=S_{\mathrm{B}}$ were perfect.

The true action $S$ is not purely bosonic, and it can be advantageous
to introduce a regularization $m_{\mathrm{reg}}$ that weakens the
effect of $\mathcal{M}$ when acting on small $\omega$. We define
diagonal matrix elements\cite{Batrouni19},

\begin{equation}
\tilde{\mathcal{M}}_{\omega,\omega}=\Delta_{\tau}\left[\frac{m_{\mathrm{reg}}^{2}+\omega_{0}^{2}+\frac{4}{\Delta_{\tau}^{2}} \sin^2 \left( \frac{2\pi\omega}{L_{\tau}}\right)}{m_{\mathrm{reg}}^{2}+\omega_{0}^{2}}\right],
\end{equation}
as the Fourier representation of the dynamical mass matrix, 
\begin{equation}
\mathcal{M}=\mathcal{F}^{-1}\tilde{\mathcal{M}}\mathcal{F}.\label{eq:dynamical_M_def}
\end{equation}

For small frequencies $\omega$ (or infinite regularization $m_{\mathrm{reg}}$)
the mass matrix is approximately constant, $\mathcal{M}\approx\Delta_{\tau}$
consistent with the scaling of fermionic forces, Eq.~(\ref{eq:SF_scaling}).
For large frequencies, $\omega\sim L_{\tau}/2$, however, a finite
regularization $m_{\mathrm{reg}}$ is irrelevant, and we find $\mathcal{M}\approx Q/\omega_{0}^{2}$.
Comparing with Eq.~(\ref{eq:dyn_bos}), the high-frequency modes
are found to behave like harmonic oscillators with an $\omega$-independent
force-scale that is again consistent with Eq.~(\ref{eq:SF_scaling}). 

The effectiveness of Fourier acceleration depends on the degree to
which a clean separation of scales can be found. Typically $\Delta_{\tau}$
will be sufficiently small such that there is a range of Fourier modes
for which $S_{\mathrm{B}}$ is the dominant contribution to the action
$S$.

Our convention for the dynamical mass matrix $\mathcal{M}$ deviates
somewhat from previous work~\cite{Batrouni19}. The present convention
aims to decouple the integration time-step $\Delta t$ from the discretization
in imaginary time $\Delta_{\tau}$, such that the two parameters may
be varied independently. In other words, one ``unit of
integration time'' should produce an approximately fixed
amount of decorrelation in the phonon field, independent of $\Delta_{\tau}$.

\begin{figure}
\includegraphics[width=1.0\columnwidth]{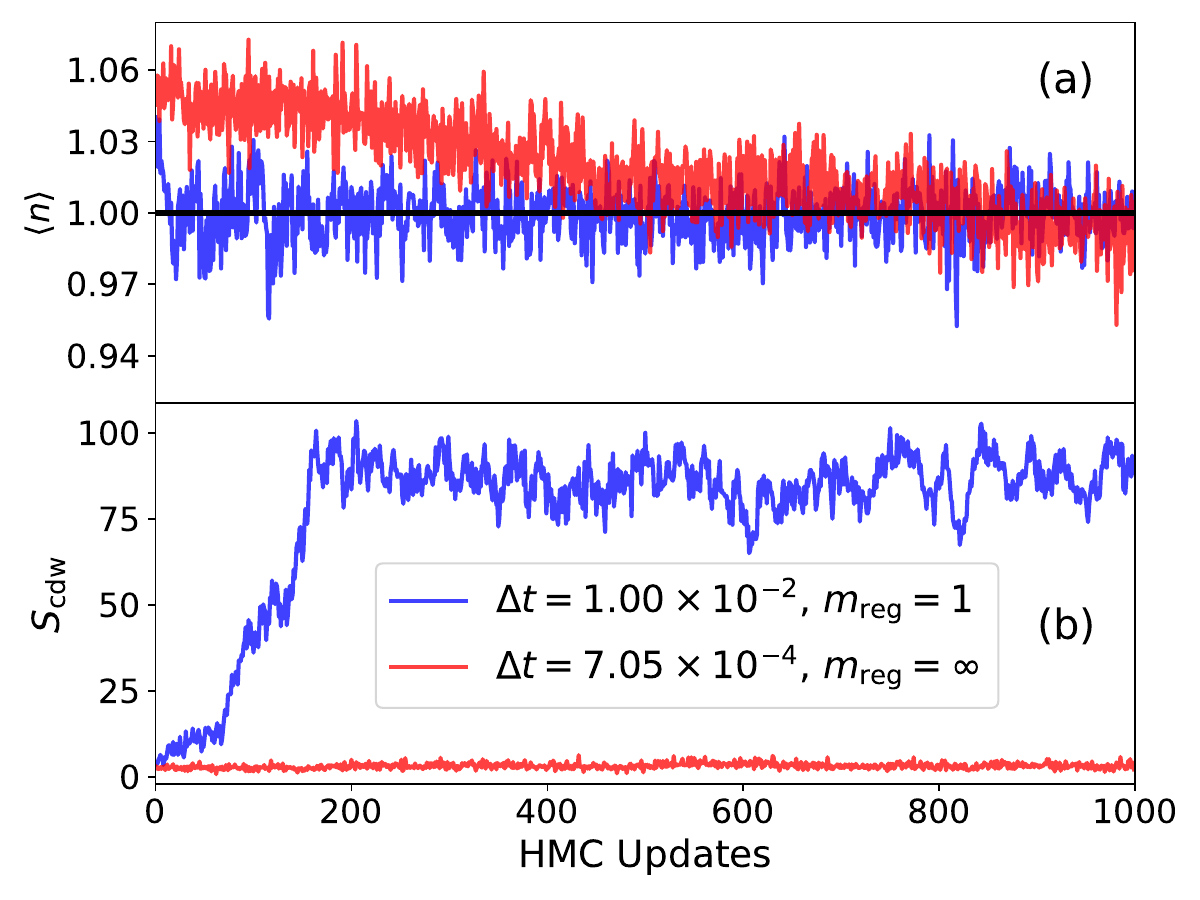}
\caption{\label{fig:fa_effect} Equilibration process for $\omega_0=1.0$, $\lambda=0.25$, $\beta=8$ and $N=256$. Panel (a) displays the time history for the
density $\left\langle n\right\rangle $. Panel (b) displays the time
history for the structure factor $S_{{\rm cdw}}$. Compares results
for two simulations started from the same initial configuration that
use different dynamical mass matrices $\mathcal{M}$. The time-steps
$\Delta t$ are chosen so that the highest frequency mode in both
simulations evolves on the same effective time-scale.}
\end{figure}

Figure~\ref{fig:fa_effect} compares the equilibration process for
two simulations of a Holstein model in the CDW phase, one using $m_{{\rm reg}}=\omega_{0}$
and $\Delta t=1\times10^{-2}$ shown in blue, the other using $m_{{\rm reg}}=\infty$
and $\Delta t=7.05\times10^{-4}$, shown in red. These $\Delta t$
have been selected such that the highest frequency dynamical mode
$\omega=L_{\tau}/2$ evolves on the same time-scales in both simulations. 

Figure~\ref{fig:fa_effect}(a) shows the time history of sampled densities
$\left\langle n\right\rangle $ for each simulation. While the measured
densities in the simulation using $m_{{\rm reg}}=\omega_{0}$ almost
immediately begin fluctuating about $\left\langle n\right\rangle =1$,
in simulations using $m_{{\rm reg}}=\infty$ the density only gradually
approaches half-filling. The discrepancy between the two simulations
is even more obvious when we look at the time series for $S_{{\rm cdw}}$
shown in Fig.~\ref{fig:fa_effect}(b). While the simulation using
$m_{{\rm reg}}=\omega_{0}$ rapidly equilibrates to CDW order in roughly
$\sim150$ updates, the $m_{{\rm reg}}=\infty$ simulation shows no
perceptible indication of thermalization towards CDW order.

\subsubsection{Time-step splitting\label{subsec:Time-step-splitting}}

A complementary strategy to handle the disparate time-scales associated
with the bosonic action is time-step splitting~\cite{Sexton92,Neal11}.
Typically, $\partial S_{\mathrm{B}}/\partial x$ is much less expensive
to evaluate than $\partial S_{\mathrm{F}}/\partial x$. One may
modify the leapfrog integration method of Eqs.~(\ref{eq:leapfrog1})--(\ref{eq:leapfrog3})
to use multiple, smaller integration timesteps $\Delta t'=\Delta t/n_{t}$
using the bosonic force alone. After taking $n_{t}$ of these sub time-steps,
a full time-step $\Delta t$ is performed using the fermionic force
alone. The final leapfrog integrator is shown in Algorithm~1, and
can be derived by a symmetric operator splitting procedure. Like the
original leapfrog algorithm, it is exactly time-reversible and symplectic.

\begin{figure}
\includegraphics[width=1.0\columnwidth]{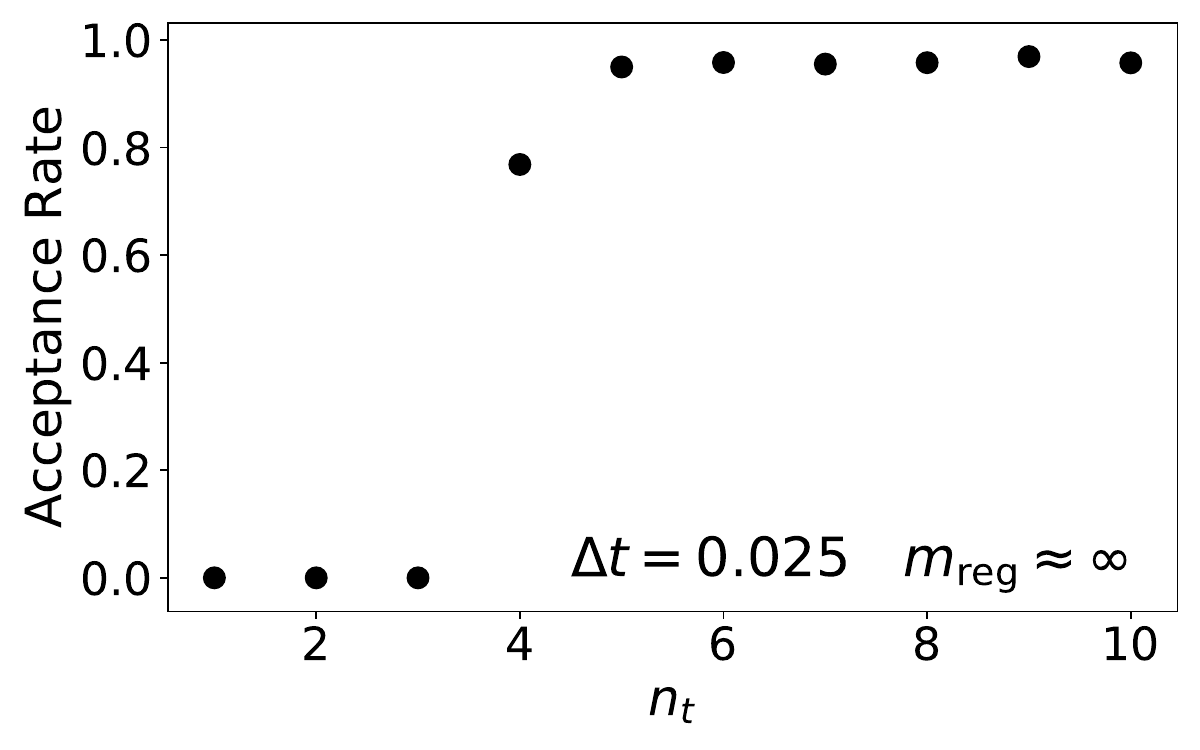}
\caption{\label{fig:split_action}HMC acceptance rate versus $n_{t}$ for $\omega_{0}=1$,
$\lambda=0.25$, $\beta=4$ and $L=16$. Note the sharp transition at $n_{t}\ge4$ sub time-steps.
For this test we disabled Fourier acceleration, effectively taking
$m_{\mathrm{reg}}=\infty$ such that $\mathcal{M}=\Delta_{\tau}$.}
\end{figure}

Figure~\ref{fig:split_action} demonstrates the practical benefit
of time-step splitting by showing how the HMC acceptance probability
varies with the number $n_{t}$ of sub-time-steps. To isolate the
impact of time-step splitting, we disabled Fourier acceleration by
effectively setting $m_{\mathrm{reg}}=\infty$. The measured acceptance
rate is zero until $n_{t}\ge4$, at which point it rapidly saturates
to a value of $\sim95\%$ once $n_{t}\ge5$. This result illustrates
a sharp stability limit: When $n_{t}<4$, the corresponding value
of $\Delta t'$ is too large to resolve the fastest Fourier modes,
$\hat{x}_{\omega=L/2}$, which causes a dynamical instability and
uncontrolled error. When $n_{t}$ increases beyond a certain point,
the corresponding values of $\Delta t'$ are sufficiently small to stabilize the $S_{\mathrm{B}}$ driven dynamics.

\subsection{Summary of an HMC update}

Algorithm~1 shows the pseudocode for one HMC trial update.

We remark that although methods of Fourier acceleration and time-step
splitting aim to solve a similar problem, they employ different mechanisms.
The dynamical mass matrix $\mathcal{M}$ of Eq.~(\ref{eq:dynamical_M_def})
was derived by analyzing a non-interacting system, and effectively
slows down the dynamics of high-frequency Fourier modes. It is effective
for handling Fourier modes for which the force contribution from $S_{\mathrm{B}}$
dominates. In contrast, time-step splitting works by focusing more
computational effort on integrating the bosonic forces, and allows
the high frequency modes to evolve on their natural, faster time-scale.
If the cost to calculate $\partial S_{\mathrm{B}}/\partial x$ were
truly negligible (relative to $\partial S_{\mathrm{F}}/\partial x$)
then we could take $n_{t}$ sufficiently large to completely resolve
the highest frequency dynamical modes arising from $S_{\mathrm{B}}$,
and Fourier acceleration could be disabled ($m_{\mathrm{reg}}\rightarrow\infty)$.
Empirically, we find a combination of the two methods to be most effective.
As such, for the rest of our benchmarks we perform HMC updates with
$\Delta t=\omega_{0}^{-1}/100$, $N_{t}=100$, $n_{t}=10$ and $m_{\mathrm{reg}}=\omega_{0}$.

\begin{algorithm}[H]
\caption{Time-step Splitting HMC Update}
\begin{algorithmic}
\State{Record initial state: $x_i$}
\State{Directly sample auxiliary field: $\Phi_\sigma \coloneqq A^T(x_i)R_\sigma$}
\State{Directly sample momentum: $p_i \coloneqq \sqrt{\mathcal{M}}R$}
\State{Calculate initial energy: $H_i \coloneqq H(x_i,p_i)$}
\For{$t \in 1 \dots N_t$}
\State{$p \coloneqq p - \frac{\Delta t}{2}\frac{\partial S_{\rm F}}{\partial x}$}
\For{$t^\prime \in 1 \dots n_t$}
\State{$p \coloneqq p - \frac{\Delta t^\prime}{2}\frac{\partial S_{\rm B}}{\partial x}$}
\State{$x \coloneqq x + \Delta t^\prime \mathcal{M}^{-1} p$}
\State{$p \coloneqq p - \frac{\Delta t^\prime}{2}\frac{\partial S_{\rm B}}{\partial x}$}
\EndFor
\State{$p \coloneqq p - \frac{\Delta t}{2}\frac{\partial S_{\rm F}}{\partial x}$}
\EndFor
\State{Calculate final energy: $H_f \coloneqq H(x_f,p_f)$}
\State{Acceptance probability: $P \coloneqq \min \left( 1 , e^{-\left( H_f - H_i \right)} \right) $}
\State{Sample $r$ uniform in $(0,1)$}
\If{$r<P$}
\State{Accept final phonon field configuration $x_f$}
\Else
\State{Revert to initial phonon field configuration $x_i$}
\EndIf
\end{algorithmic}\label{algorithm1}
\end{algorithm}

\section{Reflection and Swap Updates\label{sec:special_updates}}

Simulations of Holstein models can suffer from diverging decorrelation
times (effective ergodicity breaking) as a result of the effective phonon mediated electron-electron attraction.
The strength of this attractive interaction between electrons is approximately
parameterized by $U_{\textrm{eff}}= -\alpha^{2}/\omega_{0}^{2}=-\lambda W$
\cite{Vekic92}, where $W$ is the noninteracting bandwidth. Large dimensionless coupling $\lambda$ gives rise to ``heavy'' bipolaron physics~\cite{Esterlis19,Nosarzewski21}.
In this case, it is energetically favorable for the system to have
either 0 or 2 electrons on a site, corresponding to the phonon position
$\hat{X}$ being displaced in the positive or negative directions, respectively
(cf. Eq.~(\ref{eq:H_elph})). The energy penalty at $x_{i,\tau}=0$
roughly corresponds to the unfavorable condition of a single electron
residing on the site, and is approximately proportional to $U_{{\rm eff}}/2$.
In the context of QMC, we aim to sample fluctuations in the phonon
field $x_{i,\tau}$, with the action $S(x)$ exhibiting a strong repulsion
around $x_{i,\tau}=0$. When $\lambda$ is large, this action
barrier effectively traps the sign of the phonon field
at each site $i$.

To overcome this effective trapping, one may employ additional types
of Monte Carlo updates. We consider \emph{reflection} updates to flip
the phonon field $x_{i}\rightarrow-x_{i}$ on a single site $i$ (at
all imaginary times), and \emph{swap }updates to exchange the phonon
field $(x_{i},x_{j})\rightarrow(x_{j},x_{i})$ of neighboring sites.
Similar updates have previously been shown to be effective in DQMC
simulations of Hubbard and Holstein models~\cite{Johnston13,Scalettar91}.
A subtle difficulty arises, however, when attempting to use such global
moves in the context of fixed auxiliary fields $\Phi_{\sigma}$ (cf.
Sec.~\ref{subsec:aux}). Here we demonstrate how the introduction
of the $\Lambda$ matrix in the path integral formulation of Eq.~(\ref{eq:Z_path})
dramatically increases the acceptance rates for these global moves.

To develop intuition, we consider the single-site limit ($t_{ij}=0$)
of the Holstein model at half filling ($\mu=0$), which satisfies
an exact particle-hole symmetry. In this limit a particle-hole transformation
is realized by
\begin{equation}
\hat{X}\rightarrow-\hat{X},\quad\quad\hat{c}\rightarrow\hat{c}^{\dagger}.
\end{equation}
This transforms $\hat{n}\rightarrow1-\hat{n}$, yet leaves the Hamiltonian
$\hat{H}$ in Eq.~(\ref{eq:H_total}) invariant.

In a traditional DQMC code, the phonon field would be sampled according
to the weight $\exp\left(-S_{\mathrm{DQMC}}\right)$ appearing in
Eq.~(\ref{eq:Z_path}), where
\begin{equation}
S_{\mathrm{DQMC}}=S_{{\rm B}}-2\ln\left(e^{\beta\alpha\bar{x}/2}\det M\right)
\end{equation}
and $\bar{x}=\sum_{\tau}x_{\tau}/L_{\tau}$. In the single site limit, each
$B_{\tau}$ becomes scalar, and we can
evaluate Eq.~(\ref{eq:detM_alt}) as,
\begin{equation}
\det M=1+e^{-\Delta\tau\sum_{\tau=0}^{L_{\tau}}V_{\tau}}=1+e^{-\beta(\alpha\bar{x}-\mu)}.
\end{equation}
Taking $\mu=0$, it follows 
\begin{equation}
S_{\mathrm{DQMC}}=S_{{\rm B}}-2\ln\cosh(\beta\alpha\bar{x}/2),\label{eq:SDQMC_def}
\end{equation}
up to an irrelevant constant shift.

\begin{figure}
\includegraphics[width=1.0\columnwidth]{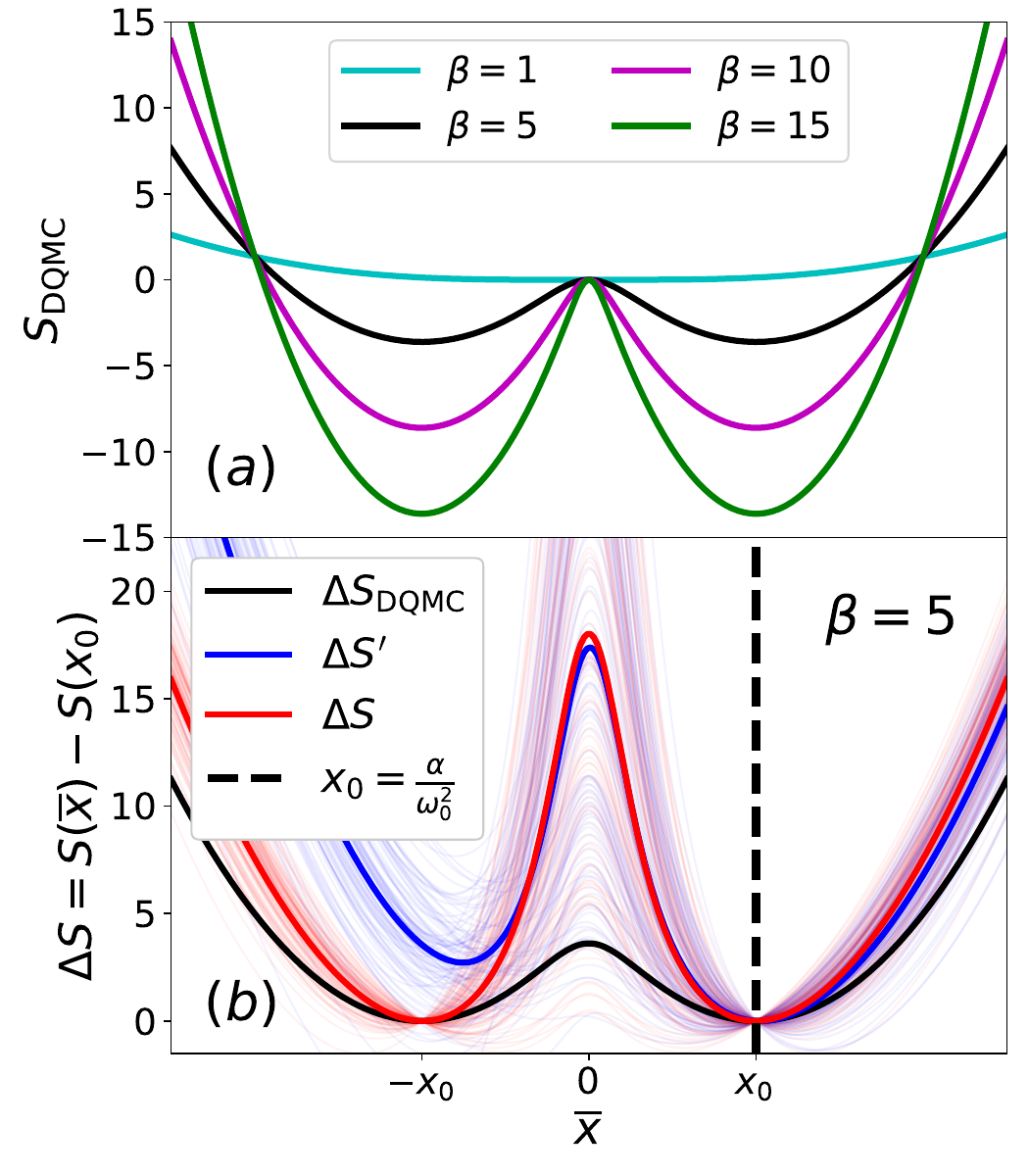}\caption{(a) $S_{\mathrm{DQMC}}(\bar{x})$ for the single-site Holstein model at half-filling
($t_{ij}=0$, $\omega_{0}=1$, $\alpha=\sqrt{2}$, $\mu=0$), plotted
as a function of the phonon field $\bar{x}$ with imaginary-time fluctuations
suppressed. (b) Change
in action under the proposed move $x_{0}\rightarrow\bar{x}$, where
$x_{0}=\alpha/\omega_{0}^{2}$. Bold blue and red lines represent the average over 100 vectors $\Phi_{\sigma}$, sampled according to Eq.~(\ref{eq:PhiSample-1}) with $\bar{x}=x_0$. With imaginary-time fluctuations suppressed,
$\Delta S$ is exactly symmetric, whereas $\Delta S'$ is not.\label{fig:DeltaS}}
\end{figure}

Let us momentarily ignore fluctuations in imaginary time,
which is justifiable at small $\omega_{0}$. By replacing $x_{\tau}\rightarrow\bar{x}$,
the bosonic action becomes $S_{\mathrm{B}}\rightarrow\beta\omega_{0}^{2}\bar{x}^{2}/2$.
Figure~\ref{fig:DeltaS}(a) plots the resulting $S_{\mathrm{DQMC}}(\bar{x})$. As the inverse temperature
$\beta$ increases, a double-well structure emerges, and the action
barrier at $\bar{x}=0$ poses a practical problem for sampling. Equation~(\ref{eq:SDQMC_def})
ensures the exact symmetry $S_{\mathrm{DQMC}}(x)=S_{\mathrm{DQMC}}(-x)$,
even in the presence of imaginary-time fluctuations, such that reflection moves would always be accepted given this choice of action.

Curiously, the $x\leftrightarrow-x$ symmetry is missing from the
action of Eq.~(\ref{eq:S_decompose}) that we actually use for sampling
the phonons. Specifically, $S(x,\Phi_{\sigma})\neq S(-x,\Phi_{\sigma})$
at fixed $\Phi_{\sigma}$. As a practical consequence, the proposal
of a global update $x\rightarrow-x$ at fixed $\Phi_{\sigma}$ may
lead to very low Monte Carlo acceptance rates, Eq.~(\ref{eq:metropolis}),
unless the action is carefully constructed.

To demonstrate how Monte Carlo acceptance rates can suffer, we consider
two $\Phi_{\sigma}$-dependent actions, $S$ and $S'$. The first
we have already defined in Eq.~(\ref{eq:S_decompose}),
\begin{equation}
S=S_{\mathrm{B}}+\frac{1}{2}\sum_{\sigma}\left|A^{-T}\Phi_{\sigma}\right|^{2},\label{eq:S_again}
\end{equation}
where $A=M\Lambda$. The second follows from Eq.~(\ref{eq:Z_path_orig}),
and would, more traditionally, be used for the Holstein model,
\begin{equation}
S'=S_{\mathrm{B}}-\beta\alpha\bar{x}+\frac{1}{2}\sum_{\sigma}\left|M^{-T}\Phi_{\sigma}\right|^{2}.\label{eq:S_prime}
\end{equation}
Both actions are statistically valid---integration over the auxiliary
fields yields the correct distribution for $x$,
\begin{equation}
\int\mathcal{D}\Phi_{\sigma}e^{-S}\propto\int\mathcal{D}\Phi_{\sigma}e^{-S'}\propto e^{-S_{\mathrm{DQMC}}}.
\end{equation}
However, the two actions produce very different acceptance rates for
global Monte Carlo moves. Figure~\ref{fig:DeltaS}(b) demonstrates
this by plotting $\Delta S$ and $\Delta S'$ for a proposed update
$x_{0}\rightarrow\bar{x}$, with imaginary-time fluctuations suppressed.
For concreteness we selected the initial condition $x_{0}=\alpha/\omega_{0}^{2}$,
but the choice does not qualitatively affect our conclusions. Each
thin curve is plotted using a different randomly sampled $\Phi_{\sigma}$,
drawn from the exponential distributions $\exp\left[-S(x_{0},\Phi_{\sigma})\right]$
or $\exp\left[-S'(x_{0},\Phi_{\sigma})\right]$ in the case of $\Delta S$
(red) or $\Delta S'$ (blue) respectively.

From Fig.~\ref{fig:DeltaS}(b), it is apparent that the action $S$ has the symmetry 
\begin{equation}
\Delta S(\bar{x}) = \Delta S(-\bar{x}).\label{eq:Delta_S_sym}
\end{equation}
This is an exact result for the single-site, adiabatic limit of the Holstein model (see Appendix~\ref{sec:S_sym}). The action
$S'$, however, has a very different qualitative behavior. Here, the proposed update $x_{0}\rightarrow -x_0$ imposes a very large action cost $\Delta S'$ for nearly all auxiliary field samples, $\Phi_\sigma$.

The qualitative difference between $\Delta S$ and $\Delta S'$ has
a profound effect on the Metropolis acceptance rate, Eq.~(\ref{eq:metropolis}),
for phonon reflections $x\rightarrow-x$. We quantify this through
numerical experiments using the single-site Holstein model at half
filling, with moderate parameters $\omega_{0}=1$, $\alpha=2$, and
$\beta=4$. If we used the full action $S_{\mathrm{DQMC}}$, the proposed
move $x\rightarrow-x$ would have a $100\%$ acceptance probability,
which follows from particle-hole symmetry, and is the ideal behavior. If the naive action $S'$
is used, the Metropolis acceptance rate for a reflection update is
only $\sim2\%$, averaged over random samples of $\Phi_{\sigma}$.
If the action $S$ is used instead of $S'$, particle-hole
symmetry is statistically restored in the sense of Eq.~(\ref{eq:Delta_S_sym}), and the acceptance
rate for reflection updates goes up to $\sim68\%$ (larger is better). We will continue to use the action $S$ throughout
the rest of this paper. Although the action $S_\mathrm{DQMC}$ yields the highest acceptance rate, it incurs a computational cost that scales cubically with system size. The action $S$ maintains a fairly high acceptance rate while retaining near-linear scaling of the overall method.
% ,  where calculating
% $\Delta S_{{\rm DQMC}}$ has a cost that scales cubically with system
% size, the procedure outlined above, that instead requires evaluating
% $\Delta S$, maintains the near linearity with system size.

\begin{figure*}
\begin{centering}
\includegraphics[width=1.75\columnwidth]{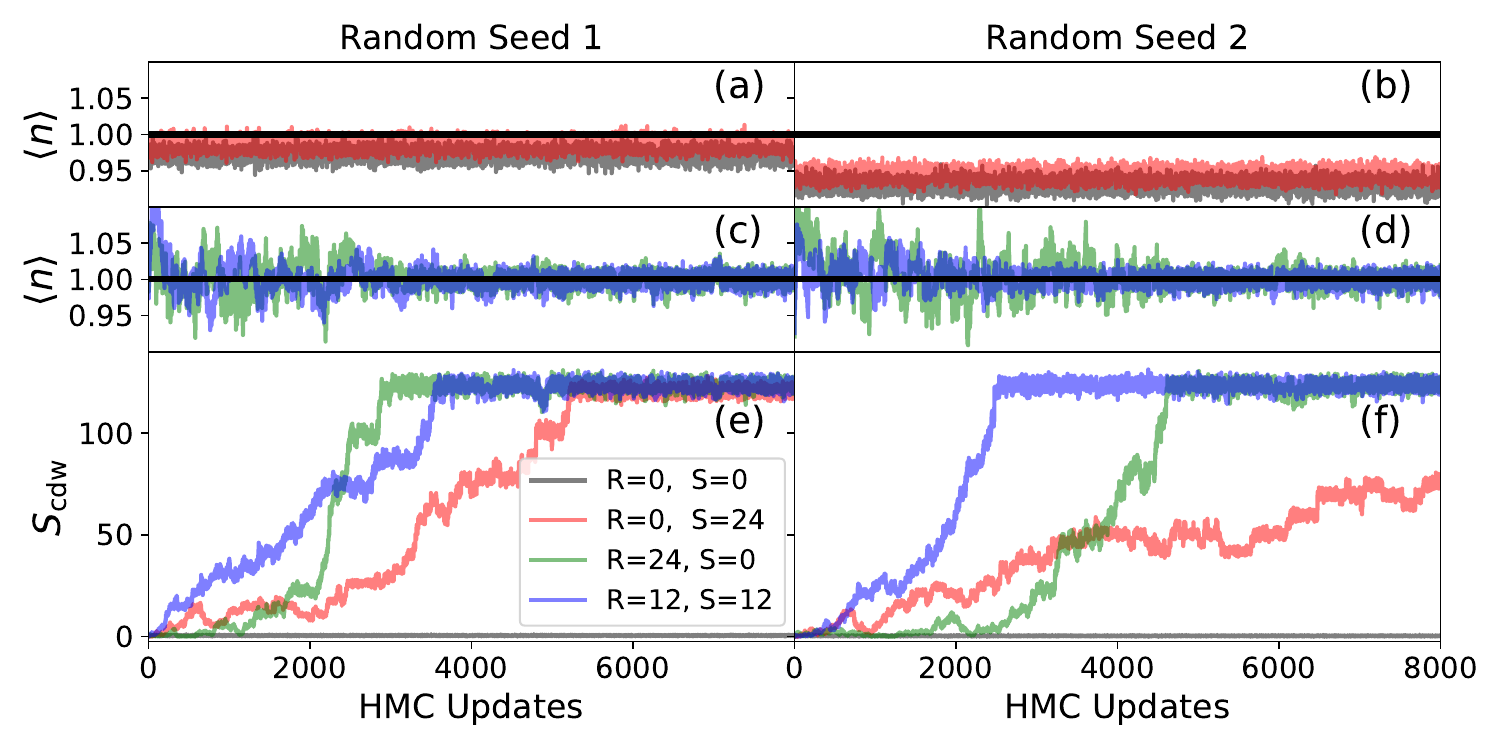}
\par\end{centering}
\caption{\label{fig:strong_coupling} Equilibration of observables with $\omega_{0}=0.1$,
$\lambda=0.6$, $\beta=9$ and $L=16$. The presence of strongly bound bipolarons leads to large sampling barriers, and simulation with HMC updates alone ($S=0, R=0$) shows a failure to equilibrate on accessible simulation scales. Including also swap updates ($S > 0, R = 0$) allows $S_\mathrm{cdw}$ to partially relax, but effective broken ergodicity is still observable from measurements of $\langle n \rangle$. Reflection updates ($R > 0$)  are crucial to realizing fast decorrelation times in all observables.}
\end{figure*}

The use of reflection and swap updates provides tremendous speed-ups
in practical studies of the Holstein model going beyond the single-site
limit. Figure~\ref{fig:strong_coupling} shows the equilibration
process for a Holstein model on a $N=16^{2}$ square lattice. We used
a relatively large coupling $\lambda=0.6$, such that on-site
action barriers are large. At inverse temperature $\beta=9$, the
system is in a robust CDW phase. We ran the same simulation twice
using two different random seeds, shown in the left and right columns.
With $\mu=0$, we know the system is at half-filling, $\left\langle n\right\rangle =1.0$.
However, in practice this correct filling fraction is only reliably
observed when reflection updates are enabled. For $S_{{\rm cdw}}$,
both reflection and swap updates help reduce decorrelation times.
In practice, using some combination of reflection and swap updates
makes sense, with reflection updates being crucial for the system
to converge properly to the correct filling.

In addition to reducing decorrelation times, reflection and swap updates
also help ameliorate a concern of ergodicity breaking~\cite{Beyl18,Beyl20,Goetz21}.
If the phonon configuration $x$ only smoothly evolves under the Hamiltonian
dynamics (Sec.~\ref{sec:hmc}) then it would be formally impossible
to cross the nodal surface where $\det M=0$, for which
the action ($S_{\mathrm{DQMC}}$ or $S$) diverges. To be sure
that we are sampling the entire space of phonon configurations, for
which $\det M$ may change sign, we should also incorporate some discontinuous
Monte Carlo updates that allow for jumps across nodal surfaces. The
reflection and swap updates proposed in this section are therefore
a good complement to pure HMC sampling.

\section{Preconditioning\label{sec:Preconditioning}}

\subsection{Preconditioner algorithm}

Each iteration of HMC requires solving the linear system in Eq.~(\ref{eq:CG}),
\begin{equation}
M^{T}Mv=b,
\end{equation}
for the unknown $v$. The required number of CG iterations to reach
a fixed level of accuracy scales approximately like the condition
number of $M$ (equivalently, the square root of the condition number
of $M^{T}M$).

Convergence can be accelerated if a good preconditioner $P$ is available.
One can solve for $u$ in
\begin{equation}
P^{-T}M^{T}MP^{-1}u=P^{-T}b\label{eq:CG_precon}
\end{equation}
and then determine $v=P^{-1}u$. This is advantageous if $MP^{-1}$
has a smaller condition number than $M$, and if $P^{-1}$ can be
efficiently to applied to a vector. In practice, each iteration
of preconditioned CG requires one matrix-vector multiplication
using $M^{T}M$, and one using $(P^T P)^{-1}$~\cite{Saad03}.

A good preconditioner frequently benefits from problem-specific insight.
For the Holstein model we make use of the fact that the $\tau$-fluctuations
in the phonon fields are damped due to the contribution to the total
action $S$ from the bosonic action $S_{{\rm B}}$ in the sampling weight $\exp(-S)$. It follows that
the imaginary-time fluctuations of the block matrices $B_{\tau}$
should be relatively small. Inspired by this, we propose a preconditioner
$P$ that retains the sparsity structure of $M$, Eq.~(\ref{eq:M_def}), but with fluctuations
in $\tau$ effectively ``averaged out.'' Specifically, we define
\begin{equation}
P=\left(\begin{array}{ccccc}
I &  &  &  & \bar{B}\\
-\bar{B} & I\\
 & -\bar{B} & I\\
 &  &  & \ddots\\
 &  &  & -\bar{B} & I
\end{array}\right),\label{eq:P_def}
\end{equation}
where
\begin{equation}
\bar{B}=\frac{1}{L_{\tau}}\sum_{\tau=0}^{L_{\tau}-1}B_{\tau}=e^{-\Delta_{\tau}\bar{V}}e^{-\Delta_{\tau}K}\label{eq:Bbar}
\end{equation}
and $\bar{V}$ is defined to satisfy
\begin{equation}
e^{-\Delta_{\tau}\bar{V}}=\frac{1}{L_{\tau}}\sum_{\tau=0}^{L_{\tau}-1}e^{-\Delta_{\tau}V_{\tau}}.\label{eq:expnVbar}
\end{equation}

This preconditioner $P$ can be
interpreted as describing a semi-classical system for which imaginary-time
fluctuations are suppressed. We emphasize, however, that our purpose with preconditioning is to solve the full Holstein model without any approximation.

Careful benchmarks of $P$ as a preconditioner to $M$ will be presented in Sec.~\ref{subsec:precon_speedup}. Here, we can briefly provide some intuition about why it should work. For small $\Delta \tau$, we have $B_\tau \approx I - \Delta_\tau (V_\tau + K)$. At this order of approximation, $\bar{V}$ becomes the imaginary time average over $V_\tau$. The bandwidth of the hopping matrix $K$ on the square lattice is $8$, whereas fluctuations in diagonal elements $(\bar{V} - V_\tau)_{ii} = \alpha (\bar{x}_i - x_{i,\tau})$ are typically order $1$ or smaller for the models considered in this work (fluctuations are controlled by $\omega_0$ when the dimensionless coupling $\lambda$ is held fixed). The relatively small magnitude of these $V_\tau$ fluctuations suggests that $P$ should be a good approximation to $M$ for the dominant part of the eigenspectrum (larger eigenvalues). We note, however, that $P$ is frequently observed to be \emph{ineffective} at capturing eigenvectors associated with the smallest eigenvalues of $M$, and this seems to be the biggest limiting factor in the utility of $P$ as a preconditioner. 

An important, but non-obvious,
property of this preconditioner is that the matrix-vector product
$P^{-1}v$ can be evaluated very efficiently. To demonstrate this, we
will first show that
the matrix $P$ becomes exactly block diagonal after an appropriate
Fourier transformation in the imaginary time $\tau$ index.

The block
structure of $M$ in Eq.~(\ref{eq:M_def}) treats $\tau=0$ as a
special case. To make all of $B_{\tau}$ appear symmetrically, we
introduce a unitary matrix,
\begin{equation}
\Theta_{\tau,\tau'}=\delta_{\tau,\tau'}e^{-\pi i\tau/L_{\tau}}.
\end{equation}
Observe that the matrix $\Theta M\Theta^{\dagger}$ has the same sparsity
structure as $M$, but a factor of $-e^{-\pi i/L}$ appears in front
of each $B_{\tau}$, and the block $B_{0}$ is no longer a special
case.

Next we may employ the discrete Fourier transformation $\mathcal{F}_{\omega,\tau}$ defined
in Eq.~(\ref{eq:F_def}). Under a combined change of basis, $M$ becomes
\begin{equation}
\tilde{M}=\mathcal{U}M\mathcal{U}^{\dagger},
\end{equation}
where
\begin{equation}
\mathcal{U}=\mathcal{F}\Theta,\label{eq:udef}
\end{equation}
is unitary, with matrix elements given by 
\begin{equation}
\mathcal{U}_{\omega,\tau}=\frac{1}{\sqrt{L_{\tau}}}e^{-\frac{2\pi i}{L_{\tau}}\tau(\omega+1/2)}.\label{eq:U_omega}
\end{equation}
By construction, the indices $\tau$ and $\omega$ range from 0 to
$L_{\tau}-1$. It is interesting to observe, however, that the 
extension of $\tau$ would naturally introduce \emph{antiperiodic} boundary conditions
($\mathcal{U}_{\omega,\tau+L_{\tau}}=-\mathcal{U}_{\omega,\tau}$),
allowing $\omega$ to be interpreted as indexing Matsubara frequencies.

Explicit calculation gives the $N\times N$ blocks of $\tilde{M}$
as
\begin{equation}
\tilde{M}_{\omega,\omega'}=\delta_{\omega,\omega'}I-e^{-i\phi_{\omega'}} \frac{\hat{B}_{\omega-\omega'}}{\sqrt{L_\tau}}, \label{eq:Mww'}
\end{equation}
where
\begin{align}
\phi_{\omega}= & \frac{2\pi}{L_{\tau}}\left(\omega+\frac{1}{2}\right)\\
\hat{B}_{\omega} =& \frac{1}{\sqrt{L_{\tau}}}\sum_{\tau=0}^{L_{\tau}-1}e^{-\frac{2\pi i}{L_{\tau}}\tau\omega}B_{\tau}.
\end{align}

We emphasize that $\tilde{M}$ is an exact representation of $M$,
but in a different basis. When the fluctuations in imaginary time are small,
$\tilde{M}$ is dominated by its diagonal blocks,
\begin{align}
\tilde{M}_{\omega,\omega}= & I-e^{-i\phi_{\omega}}\bar{B},\label{eq:Mww}
\end{align}
where $\bar{B} = \hat{B}_0 / \sqrt{L_\tau}$ coincides with Eq.~(\ref{eq:Bbar}).

We may \emph{define} the preconditioner to be block diagonal in the
Fourier basis,
\begin{equation}
\tilde{P}=\textrm{diag}(\tilde{M}).\label{eq:P_block}
\end{equation}
Transforming back to the original basis,
\begin{equation}
P=\mathcal{U}^{\dagger}\tilde{P}\mathcal{U},
\end{equation}
makes contact with the equivalent definition in Eq.~(\ref{eq:P_def}).

To apply the preconditioner to a vector $v$, we must evaluate
\begin{equation}
P^{-1}v=\mathcal{U}^{\dagger}\tilde{P}^{-1}\mathcal{U}v\label{eq:P_change_basis}
\end{equation}
The action of $\mathcal{U}$ and $\mathcal{U}^{\dagger}$ can be efficiently
implemented using a fast Fourier transform (FFT). Because $\tilde{P}$
is block diagonal, its inverse is also block diagonal,
\begin{equation}
\tilde{P}_{\omega,\omega'}^{-1}=\delta_{\omega,\omega'}\tilde{M}_{\omega,\omega}^{-1}.
\end{equation}
Therefore, applying $\tilde{P}^{-1}$ to a $\left(N\times L_{\tau}\right)$-dimensional
vector $\hat{v}=\mathcal{U}v$ is equivalent to applying each of the
$\tilde{M}_{\omega,\omega}^{-1}$ blocks to the corresponding $N$-dimensional
sub-vector $\hat{v}_{\omega}$. In Appendix~\ref{sec:precond_impl}
we describe how the kernel polynomial method (KPM)~\cite{Weisse06}
can be used to carry out efficiently these matrix-vector multiplications.
The key idea is to approximate each of $\tilde{M}_{\omega,\omega}^{-1}$
using a numerically stable Chebyshev series expansion in powers of
the matrix $\bar{B}$.

\subsection{Preconditioner speed-up \label{subsec:precon_speedup}}

Here we present results that demonstrate the utility of our preconditioner
$P$, while also providing insight into the scaling of HMC with both
system size $N$ and inverse temperature $\beta$. The overwhelming
computational cost in HMC is repeatedly solving the linear system
Eq.~(\ref{eq:CG}) for varying realizations of the phonon field $x_{i,\tau}$.
If the number of CG iterations required to find a solution is independent
of $N$, then the total simulation cost would scale near linearly
with $N$. 

In all cases, we terminate the CG iterations when the relative magnitude
of the residual error, 
\begin{equation}
\epsilon=\left|b-M^{T}Mv\right|/\left|b\right|
\end{equation}
becomes less than a threshold value $\epsilon_{{\rm max}}$. When
calculating $\Delta S_{{\rm F}}$ in Eq.~(\ref{eq:Delta_S_F}) to
 accept or reject a Monte Carlo update, we use $\epsilon_{{\rm max}}=10^{-10}$.
When calculating $\partial S_{\mathrm{F}}/\partial x$ in Eq.~(\ref{eq:dSF_dx})
we use $\epsilon_{{\rm max}}=10^{-5}$.

\begin{figure}
\includegraphics[width=1.0\columnwidth]{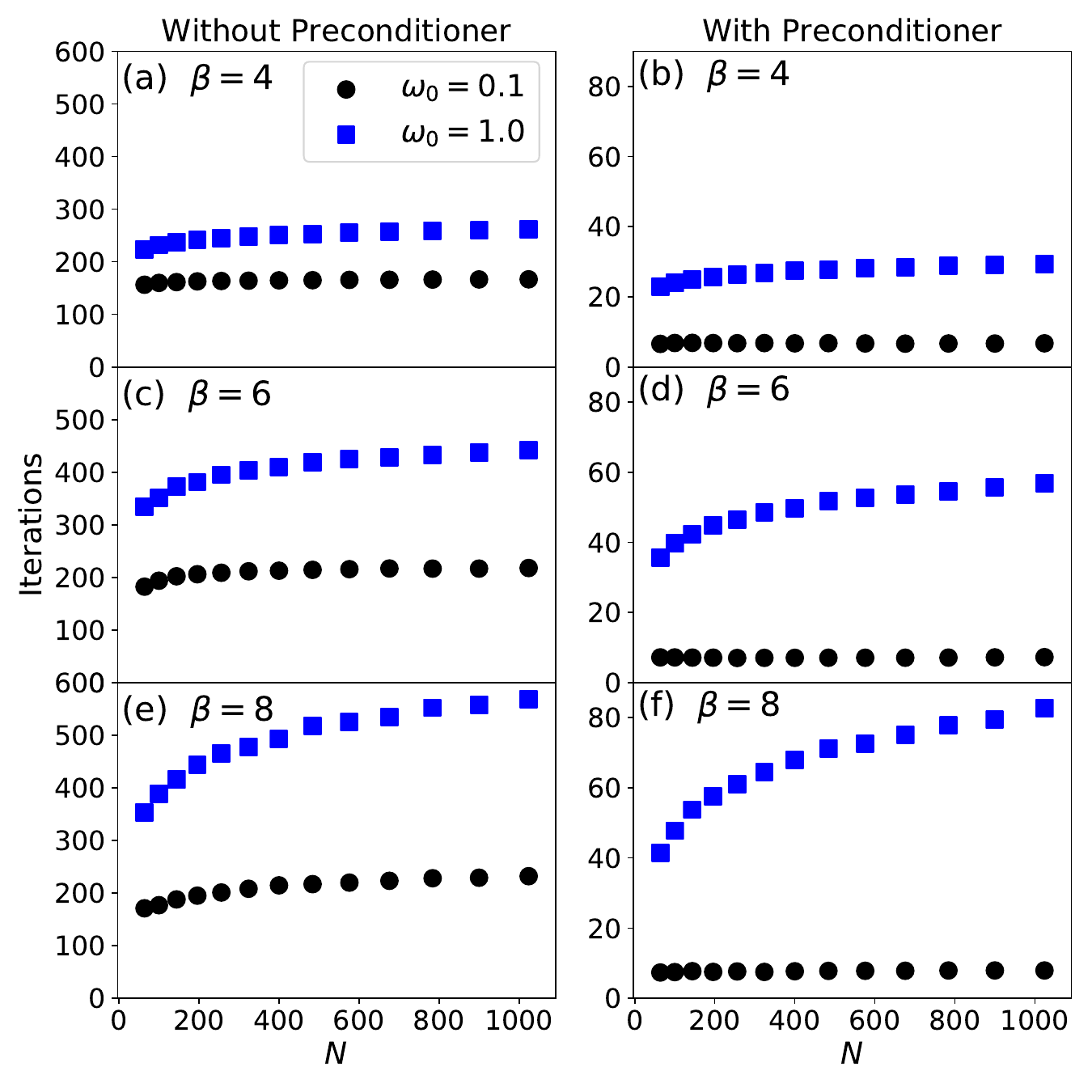}
\caption{\label{fig:iters_vs_N}Average CG iteration count as a function of
system size $N$ for $\lambda=0.25$. Comparing the left and right
columns, the preconditioner significantly reduces the iteration count.}
\end{figure}

We benchmark using Holstein models of various systems sizes at two
phonon frequencies $\omega_{0}=0.1$ and $\omega_{0}=1.0$, both with
dimensionless coupling $\lambda=0.25$. Figure~\ref{fig:iters_vs_N} shows the average iteration count as
a function of the number of lattice sites, $N$. For all temperatures
and lattice sizes, the $\omega_{0}=0.1$ simulations require fewer
CG iterations than comparable $\omega_{0}=1.0$ simulations.
%In the latter case, the condition number of $M$ is observed to be larger.
For $\omega_{0}=0.1$ without the preconditioner, the iteration count
only weakly depends on system size. However, with the preconditioner
the iteration count becomes nearly independent of system size, and
is decreased by more than a factor of $20$. For $\omega_{0}=1.0$,
the growth of CG iteration count as a function of system size remains
sub-linear. Introducing the preconditioner does not change the qualitative
structure of this dependence, but still reduces
the iteration count by more than a factor of $5$ in all cases.

\begin{figure}
\includegraphics[width=1.0\columnwidth]{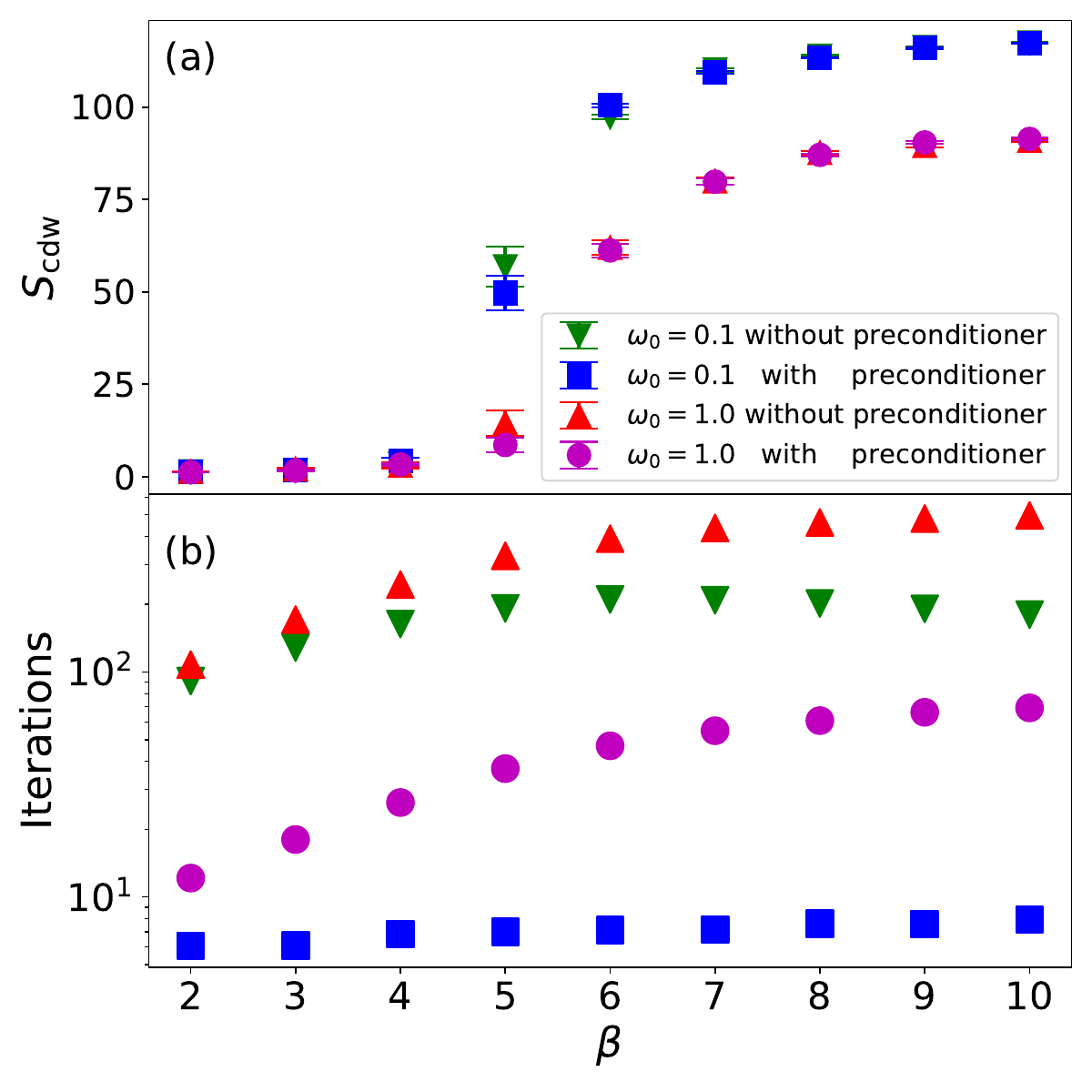}
\caption{\label{fig:iters_vs_beta}$S_{{\rm cdw}}$ and average CG iteration
count as functions of $\beta$ for $\lambda=0.25$ and $N=16^{2}$.
For both $\omega_{0}=0.1$ and $\omega_0=1.0$, the system goes through a CDW transition
as the temperature is lowered. In the case of $\omega=1.0$ the known
transition temperature is approximately $\beta_{{\rm cdw}}\approx6$.}

\end{figure}

The dependence of iteration count on $\beta$ is further explored in Fig.~\ref{fig:iters_vs_beta}.
For both $\omega_{0}=0.1$ and $\omega_{0}=1.0$, we observe a sharp
jump in the order parameter $S_{{\rm cdw}}$ as the temperature is lowered,
indicating that both systems order into a CDW phase. In the lower
panel we see the average iteration count versus $\beta$. Both with
and without the preconditioner, in the case of $\omega_{0}=1.0$ the
iteration count increases monotonically with $\beta$. Simulations
with $\omega_{0}=0.1$ have two qualitatively different behaviors:
with preconditioning, the iteration count is relatively flat, but
without preconditioning, the iteration count has a local maximum near
where we would estimate the transition temperature to be based on
$S_{{\rm cdw}}$.

The preconditioner significantly reduces the average iteration count
for both $\omega_{0}=0.1$ and 1, but the effect is more pronounced
for smaller $\omega_{0}$, where imaginary time fluctuations are smaller.
In the adiabatic limit, corresponding to the atomic mass going to
infinity, the fluctuations
in $\tau$ would vanish, and the preconditioner would become perfect.
The adiabatic limit can equivalently be reached by sending the phonon
frequency to zero $\left(\omega_0\rightarrow0\right)$ while holding
$\lambda$ fixed.

The \emph{practical} benefit of preconditioning depends strongly on
the numerical cost $C_{P}$ to apply the preconditioner $P^{-1}$
to a vector. The natural reference scale is $C_{M}$, the cost to
apply the unpreconditioned matrix $M$ to a vector. In our implementation,
we measure $(C_{P}+C_{M})/C_{M}\approx4$, approximately independent
of model details (see Sec.~\ref{subsec:precond_scaling} for a theoretical
analysis). At $\omega_{0}=0.1$, preconditioning reduces the iteration
count by about a factor of 20, yielding an effective speedup of order
$20/4=5$.

Wall-clock times for the simulation results reported in Figs.~\ref{fig:iters_vs_N}~and~\ref{fig:iters_vs_beta} can be found in
Appendix~\ref{sec:wall_clock_time}. The results confirm that the computational cost scales near-linearly in both system size and inverse temperature $\beta$. Furthermore, the speedups due to preconditioning are very close to the estimates given above.

\section{Stochastic Measurements with FFT acceleration\label{sec:measurements}}

In a traditional determinant QMC code, measurements of the Green function
are obtained by explicit construction of the matrix $M^{-1}$. This cubic-scaling cost can be avoided by using stochastic techniques to estimate
individual matrix elements. We review these methods, and then demonstrate
how to efficiently average Green function elements over all space and imaginary
times by using the FFT algorithm. Finally, we will introduce
a strategy to reduce the relatively large stochastic errors that appear
when forming stochastic estimates of multiple-point correlation functions.

\subsection{Measurements in QMC}

A fundamental observable in QMC simulation is the time-ordered, single-particle
Green function,
\begin{equation}
\mathcal{G}_{i,j}\left(\tau\right)=\begin{cases}
\langle\hat{c}_{i}(\tau)\hat{c}_{j}^{\dagger}(0)\rangle, & 0\leq\tau<\beta\\
-\langle\hat{c}_{j}^{\dagger}(0)\hat{c}_{i}(\tau)\rangle, & -\beta\leq\tau<0
\end{cases}.
\end{equation}
where $\hat{c}_{i}(\tau)\equiv e^{\tau H}\hat{c}_{i}e^{-\tau H}$
denotes evolution of the electron annihilation operator in continuous imaginary time $\tau$. Multi-point
correlation functions can be expressed as sums of products of single-particle
Green functions via Wick's theorem~\cite{Gubernatis16,Assaad08}.
Given an equilibrium sample of the phonon field, the matrix $G=M^{-1}$
provides an unbiased estimate of the Green function, 
\begin{equation}
\mathcal{G}_{i,j}\left(\tau\right)\approx G_{(i,l),(j,l')},\label{eq:Green_approx}
\end{equation}
where $\tau=\Delta_{\tau}\cdot(l-l')$ satisfies $-\beta<\tau<\beta$.
In what follows we will revert to using the symbol $\tau=0,1\dots,(L_{\tau}-1)$
as a matrix index instead of a continuous imaginary time.

\subsection{Stochastic approximation of the Green function}

In a traditional determinant QMC code, one would explicitly calculate
the full matrix $G=M^{-1}$ at a cost that scales cubically in system
size. To reduce this cost, we instead employ the unbiased stochastic
estimator 
\begin{equation}
G\approx(G\xi)\xi^{T},\label{eq:Hutch_approx}
\end{equation}
for a random vector $\xi$ with components that satisfy $\langle\xi_{\boldsymbol{i}}\rangle=0$
and $\langle\xi_{\boldsymbol{i}}\xi_{\boldsymbol{j}}\rangle=\delta_{\boldsymbol{i},\boldsymbol{j}}$.
For example, each component $\xi_{\boldsymbol{i}}$ may be sampled
from a Gaussian distribution, or uniformly from $\{\pm1\}$. The bold symbol $\boldsymbol{i}$ represents a combined site
and imaginary-time index, $(i,\tau$). Equation~(\ref{eq:Hutch_approx}) may be viewed as a generalization
the Hutchinson trace estimator $\mathrm{Tr}\,G\approx\xi^{\dagger}G\xi$
\cite{Hutchinson90}. Various strategies are possible to reduce
the stochastic error~\cite{Tang12,Wang18a}.

The vector $v=G\xi$ can be calculated iteratively at a cost that
scales near-linearly with system size. For example, one may solve
the linear system $M^{T}Mv=M^{T}\xi$ using CG with preconditioning
(cf. Sec.~\ref{sec:Preconditioning}). 

Once $G\xi$ is known, individual matrix elements can be efficiently
approximated,
\begin{equation}
G_{\boldsymbol{i},\boldsymbol{j}}\approx\left(G\xi\right)_{\boldsymbol{i}}\xi_{\boldsymbol{j}}.\label{eq:G_approx}
\end{equation}
For products of Green functions elements, we may use,
\begin{equation}
G_{\boldsymbol{i},\boldsymbol{j}}G_{\boldsymbol{k},\boldsymbol{l}}\approx\left(G\xi\right)_{\boldsymbol{i}}\xi_{\boldsymbol{j}}\left(G\xi'\right)_{\boldsymbol{k}}\xi'_{\boldsymbol{l}}.\label{eq:GG_approx}
\end{equation}
This product of estimators remains an unbiased estimator provided
that the random vectors $\xi$ and $\xi'$ are mutually independent.

\begin{figure*}
\includegraphics[width=1.75\columnwidth]{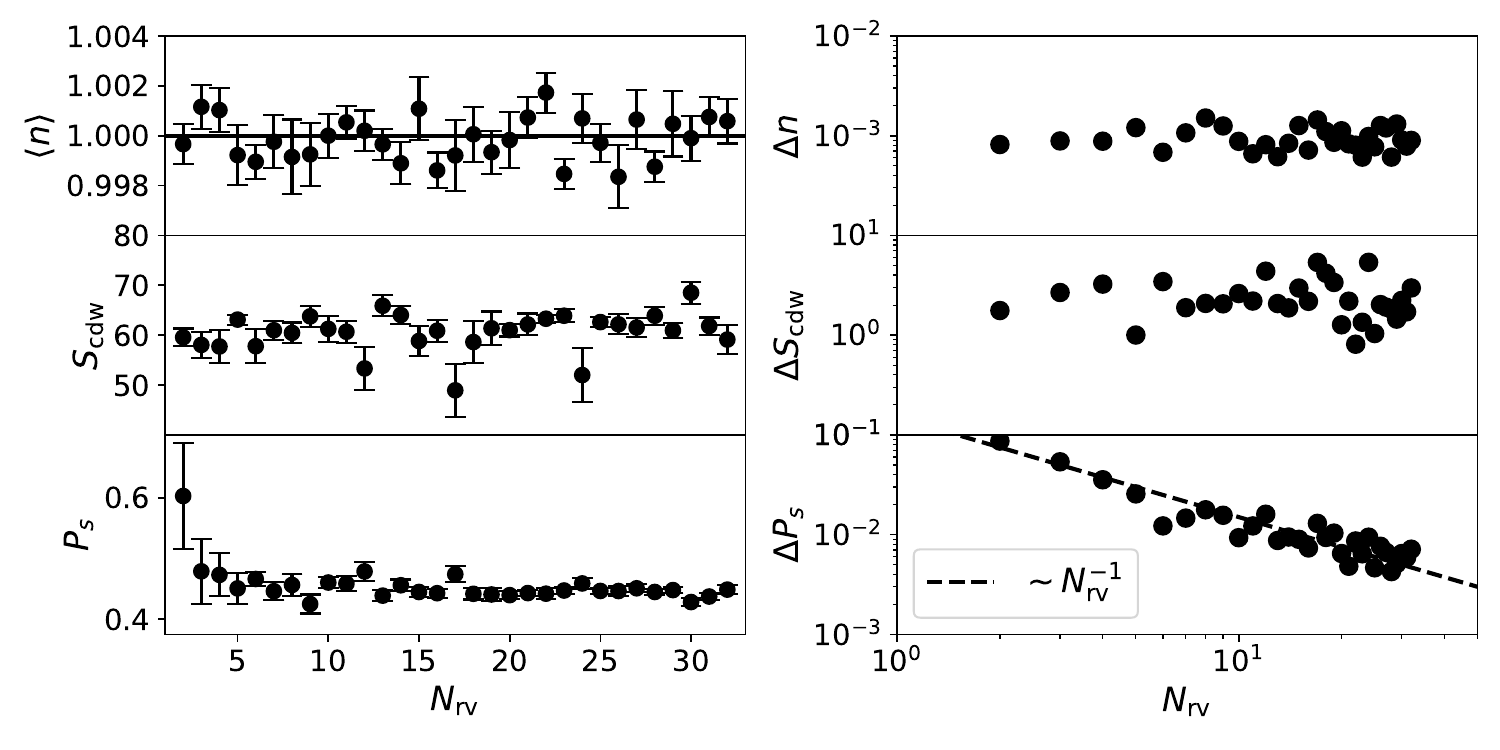}
\caption{\label{fig:error_vs_Nrv}Left: QMC measurements, employing stochastic
Green function estimation with $N_{{\rm rv}}$ random vectors. Right:
Error for each measured quantity. Simulations were performed using
$\omega=1$, $\lambda=0.25$, $\beta=6$, $L=16$, $N_{{\rm therm}}=1000$
and $N_{{\rm sim}}=2000$.}
\end{figure*}

\subsection{Averaging over space and imaginary time using FFTs}

To improve the quality of statistical estimates, it is frequently
desirable to average Green function elements over all space and imaginary-time,
\begin{equation}
\mathcal{G}_{\boldsymbol{\Delta}}\approx\frac{1}{\mathcal{N}}\sum_{\boldsymbol{i}}G_{(\boldsymbol{i}+\boldsymbol{\Delta}),\boldsymbol{i}}^{\mathrm{ext}},\label{eq:G_avg-1}
\end{equation}
where $\mathcal{N}=NL_{\tau}$. The symbol $\boldsymbol{\Delta}$
indicates a displacement in both position and imaginary-time. The
matrix $G^{\mathrm{ext}}$ will be defined below as an extension of
$G$ that accounts for antiperiodicity of imaginary time. Using direct
summation, the total cost to calculate $\mathcal{G}_{\boldsymbol{\Delta}}$
for every possible displacement $\boldsymbol{\Delta}$ would scale
like $\mathcal{O}\left(\mathcal{N}^{2}\right)$. However, we will
describe a method using FFTs that reduces the cost to approximately
$\mathcal{O}\left(\mathcal{N}\ln\mathcal{N}\right)$.

Consider a finite, $D$-dimensional lattice with periodic boundary
conditions. For a Bravais lattice, each site can be labeled by integer
coordinates, $0\leq n_{d}<L_{d}$, where $L_d$ is the linear system size for dimension $d$. The combined index $\boldsymbol{i}=\left(n_{1},\dots,n_{D},\tau\right)$
can then be interpreted as integer coordinates for both space and
imaginary-time; the index $(\boldsymbol{i}+\boldsymbol{\Delta})$
can be interpreted as a displacement of all $(D+1)$ coordinates.
We must be careful, however, with boundary conditions. The Green function
is \emph{antiperiodic} in continuous imaginary time, $\mathcal{G}_{i,j}\left(\tau+\beta\right)=-\mathcal{G}_{i,j}\left(\tau\right)$.
To encode this antiperiodicity in matrix elements, we define
\begin{equation}
G^{\mathrm{ext}}=QGQ^{T}=\left[\begin{array}{cc}
G & -G\\
-G & G
\end{array}\right],\label{eq:G_def-1}
\end{equation}
where 
\begin{equation}
Q=\left[\begin{array}{c}
I\\
-I
\end{array}\right].
\end{equation}
The extended matrix $G^{\mathrm{ext}}$ effectively doubles the range
of the imaginary time index, $0\leq\tau<2L_{\tau}$, such that space
\emph{and }imaginary time indices become periodic,
\begin{equation}
n_{d}+L_{d}\equiv n_{d},\quad\quad\tau+2L_{\tau}\equiv\tau.
\end{equation}

Using Eq.~(\ref{eq:Hutch_approx}), we obtain a stochastic approximation
for the time averaged Green function elements,
\begin{align}
\mathcal{G}_{\boldsymbol{\Delta}} & \approx\frac{1}{\mathcal{N}}\sum_{\boldsymbol{i}}\left(QG\xi\xi^{T}Q^{T}\right)_{(\boldsymbol{i}+\boldsymbol{\Delta}),\boldsymbol{i}}\nonumber \\
 & =\frac{1}{\mathcal{N}}\sum_{\boldsymbol{i}}a_{\boldsymbol{i}}b_{\boldsymbol{i+\Delta}},
\end{align}
involving the vectors
\begin{equation}
a=Q\xi=\left[\begin{array}{c}
\xi\\
-\xi
\end{array}\right],\quad\quad b=QG\xi=\left[\begin{array}{c}
G\xi\\
-G\xi
\end{array}\right].
\end{equation}
This can be written,
\begin{equation}
\mathcal{G}_{\boldsymbol{\Delta}}\approx\frac{1}{\mathcal{N}}(a\star b)_{\boldsymbol{\Delta}},
\end{equation}
where $a\star b$ denotes the circular cross-correlation. Like the convolution operation,
it can be expressed using ordinary multiplication in Fourier space,
\begin{equation}
(a\star b)_{\boldsymbol{\Delta}}=\mathcal{F}^{-1}\{\mathcal{F}[a]^{*}\mathcal{F}[b]\}_{\boldsymbol{\Delta}}.\label{eq:cross_corr}
\end{equation}
Here, $\mathcal{F}$ denotes the $(D+1)$-dimensional discrete Fourier
transform. This formulation allows using the FFT algorithm to estimate $\mathcal{G}_{\boldsymbol{\Delta}}$ at near-linear scaling cost.

In the QMC context, Wick's theorem ensures that multi-point correlation
functions can always be reduced to products of ordinary Green functions.
The latter can be estimated using a product of independent stochastic
approximations, as in Eq.~(\ref{eq:GG_approx}). Here, again, we
can accelerate space and imaginary-time averages using FFTs. In the
case of 4-point measurements, Wick's theorem produces three types
of Green function products. The first is,

\begin{align}
\sum_{\boldsymbol{i}}G_{\boldsymbol{i}+\boldsymbol{\Delta},\boldsymbol{i}}G_{\boldsymbol{i}+\boldsymbol{\Delta},\boldsymbol{i}} & \approx\sum_{\boldsymbol{i}}\left(G\xi\xi^{T}\right)_{\boldsymbol{i}+\boldsymbol{\Delta},\boldsymbol{i}}\left(G\xi'\xi'^{T}\right)_{\boldsymbol{i}+\boldsymbol{\Delta},\boldsymbol{i}}\nonumber \\
 & =\sum_{\boldsymbol{i}}\left[\xi_{\boldsymbol{i}}\xi'_{\boldsymbol{i}}\right]\left[\left(G\xi\right)_{\boldsymbol{i}+\boldsymbol{\Delta}}\left(G\xi'\right)_{\boldsymbol{i}+\boldsymbol{\Delta}}\right]
\end{align}
which is again recognized as a cross correlation $\star$. This can
be expressed compactly by introducing $\odot$ to denote element-wise
multiplication of vectors,
\begin{equation}
\sum_{\boldsymbol{i}}G_{\boldsymbol{i}+\boldsymbol{\Delta},\boldsymbol{i}}G_{\boldsymbol{i}+\boldsymbol{\Delta},\boldsymbol{i}}\approx\left[\left(\xi\odot\xi'\right)\star\left(G\xi\odot G\xi'\right)\right]_{\boldsymbol{\Delta}},
\end{equation}
The other two averages that appear for 4-point measures can be expressed
similarly,
\begin{align}
\sum_{\boldsymbol{i}}G_{\boldsymbol{i}+\boldsymbol{\Delta},\boldsymbol{i}+\boldsymbol{\Delta}}G_{\boldsymbol{i},\boldsymbol{i}} & \approx\left[\left(\xi\odot G\xi\right)\star\left(\xi'\odot G\xi'\right)\right]_{\boldsymbol{\Delta}}\\
\sum_{\boldsymbol{i}}G_{\boldsymbol{i}+\boldsymbol{\Delta},\boldsymbol{i}}G_{\boldsymbol{i},\boldsymbol{i}+\boldsymbol{\Delta}} & \approx\left[\left(\xi\odot G\xi'\right)\star\left(\xi'\odot G\xi\right)\right]_{\boldsymbol{\Delta}}.
\end{align}
\begin{figure}

\includegraphics[width=0.95\columnwidth]{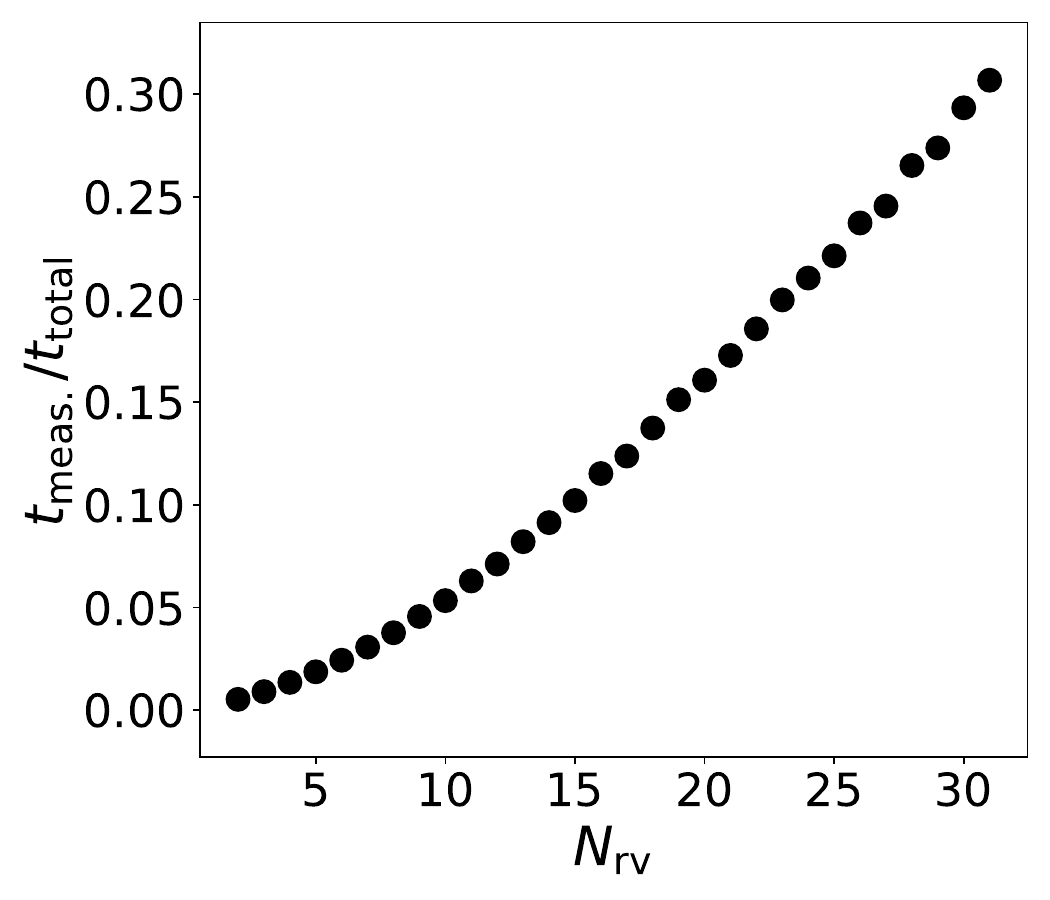}
\caption{\label{fig:meas_time_vs_Nrv}Wall-clock time spent taking measurements $t_{{\rm meas}}$
relative to the total run-time $t_{{\rm total}}$ as a function of
$N_{{\rm rv}}$. Simulation parameters are the same as in Fig.~\ref{fig:error_vs_Nrv}.}
\end{figure}

\subsection{Reducing stochastic error in multi-point correlation function estimates}

To reduce the stochastic error in Eq.~(\ref{eq:G_approx}), we may
average over a collection of random vectors,  $[\xi_{1},\dots\xi_{N_{\mathrm{rv}}}]$,
\begin{equation}
G_{\boldsymbol{i},\boldsymbol{j}}\approx\frac{1}{N_{\mathrm{rv}}}\sum_{n=1}^{N_{\mathrm{rv}}}\left(G\xi_{n}\xi_{n}^{\dagger}\right)_{\boldsymbol{i},\boldsymbol{j}}.\label{eq:G_approx_better}
\end{equation}
A similar strategy could be used to replace Eq.~(\ref{eq:GG_approx})
with an average over $N_{\mathrm{rv}}$ independent estimates.

A significant reduction in error is possible by averaging over all ${N_{{\rm rv}} \choose 2}=N_{\mathrm{rv}}(N_{\mathrm{rv}}-1) /2 $ pairs of random vectors,
\begin{equation}
G_{\boldsymbol{i},\boldsymbol{j}}G_{\boldsymbol{k},\boldsymbol{l}}\approx{N_{{\rm rv}} \choose 2}^{-1}\sum_{n < m}\left(G\xi_{n}\xi_{n}^{\dagger}\right)_{\boldsymbol{i},\boldsymbol{j}}\left(G\xi{}_{m}\xi{}_{m}^{\dagger}\right)_{\boldsymbol{k},\boldsymbol{l}}.\label{eq:GG_approx_better}
\end{equation}
This improved estimator is an average of unbiased estimators and therefore
remains unbiased. Furthermore, if $N_{\mathrm{rv}}$ is much smaller
than the vector dimension $\mathcal{N}$, then these ${N_{{\rm rv}} \choose 2}\approx N_{\mathrm{rv}}^{2}$
estimates are approximately mutually independent. It follows that
the stochastic error in Eq.~(\ref{eq:GG_approx_better}) decays approximately
like $N_{\mathrm{rv}}^{-1}$. This scheme is advantageous because,
for moderate $N_{\mathrm{rv}}$, the dominant computational cost is
calculating the $N_{\mathrm{rv}}$ matrix-vector products $[G\xi_{1},\dots G\xi_{N_{\mathrm{rv}}}]$.
There remains the task of evaluating the sum over all pairs $n\neq m$.
For each pair, we must evaluate cross-correlations as in Eq.~(\ref{eq:cross_corr}),
but the required FFTs are relatively fast.

Figure~\ref{fig:error_vs_Nrv} demonstrates how the improved stochastic
approximator in Eq.~(\ref{eq:GG_approx_better}) can significantly
reduce error bars for certain observables in QMC simulation. Measurements
and corresponding estimated errors are plotted as a function of $N_{{\rm rv}}$.
For the observables $\left\langle n\right\rangle $ and $S_{\mathrm{cdw}}$,
the error appears largely independent $N_{{\rm rv}}$; in these two
cases, the dominant source of statistical error seems to be limited
by the effective number of independent phonon configurations sampled.

For the observable $P_{s}$, however, we find the error $\Delta P_{s}$
to depend strongly on the quality of the stochastic estimator, controlled
by $N_{\mathrm{rv}}$. The observed scaling $\Delta P_{s}\sim N_{\mathrm{rv}}^{-1}$
matches the theoretical expectation for stochastic error in Eq.~(\ref{eq:GG_approx_better}).
This indicates that the stochastic measurements are the primary source
of error in $P_{{\rm s}}$.

It is also important to consider the relative computational cost of
measurements as $N_{{\rm rv}}$ increases. Figure~\ref{fig:meas_time_vs_Nrv}
plots the time spent making measurements $t_{{\rm meas}}$, relative
to the total simulation time $t_{{\rm total}}$, versus $N_{{\rm rv}}$.
Even at the maximum value of $N_{{\rm rv}}=32$ tested, the time spent
making measurements is significantly less than half the total run-time.
The fact that the $t_{{\rm meas}}/t_{{\rm total}}$ grows linearly
at large $N_{{\rm rv}}$ indicates that calculating the matrix-vector
products $G \xi_n$ is the dominant
computational cost in the measurement process. The curvature at small
$N_{{\rm rv}}$ is a result of $t_{{\rm total}}$ including the overhead
time spent equilibrating the system before measurements begin. A practical
limitation on $N_{{\rm rv}}$ may be memory usage, since Eq.~(\ref{eq:GG_approx_better})
requires that all vectors $[\xi_{1},\dots\xi_{N_{\mathrm{rv}}}]$
and $[G\xi_{1},\dots G\xi_{N_{\mathrm{rv}}}]$ be stored simultaneously. 

Although $N_{\mathrm{rv}}$ appears to have little impact on some
observables, it seems reasonable to set $N_{{\rm rv}}\gtrsim10$ in
most cases, given the negligible computational costs.

\section{Conclusion\label{sec:conclusion}}

This paper introduces a set of algorithms that collectively enable
highly scalable, finite temperature simulations of electron-phonon models such as the Holstein and SSH models. Traditionally, such studies would be performed using DQMC, but that approach is limited in two important respects.

First, with a computational cost that scales cubically with system
size, DQMC simulations of the Holstein model have been restricted
to lattices of no more than a few hundred sites. As a result, DQMC
studies of the Holstein model have typically been confined to relatively
simple geometries in one or two dimensions. In the HMC approach explored in this paper,
we replace each Fermion determinant $\det M(x)$ that appears in DQMC with a
Gaussian integral over a newly introduced auxiliary field $\Phi_\sigma$
(Sec.~\ref{subsec:aux}). This field must be multiplied by the inverse matrix $M^{-1}(x)$; for this, we use the iterative conjugate gradient (CG) method, with a computational cost that scales near-linearly with system size. As a result, it becomes possible to simulate lattice sizes
a full order of magnitude larger than is possible with DQMC. We
accelerate CG convergence by introducing a preconditioner
$P$ that retains the structure of $M(x)$, but discards fluctuations in imaginary time  (Sec.~\ref{sec:Preconditioning}).
These advances open the door to studying both more complicated multi-band
models in two dimensions, as well as three dimensional systems.

Second, DQMC simulations rely on a local updating scheme that results
in long autocorrelation times that increase with decreasing phonon
frequency. This has restricted DQMC simulations to systems where the
phonon energy is comparable to the hopping amplitude, $\omega_{0}\sim t$.
However, in most real materials the relative phonon energy is much
smaller, $\omega_{0}\ll t$. We address this limitation by using HMC
to update efficiently the entire phonon field simultaneously. To
do so, we employ a Hamiltonian dynamics with a carefully defined dynamical
mass matrix that slows down the modes with highest frequency in imaginary time,  
which counteracts stiffness in the bosonic action $S_{{\rm B}}$ (Sec.~\ref{subsec:mass_matrix}).
Additionally, we employ a time-step splitting algorithm (Sec.~\ref{subsec:Time-step-splitting})
that uses a smaller time-step to integrate the bosonic forces $- \partial S_{{\rm F}}/\partial x$, relative to the time-step for the fermionic forces.
As a result, we are able to simulate efficiently electron-phonon models with small phonon frequencies, which are of greatest physical relevance for real materials.

At moderate to strong electron-phonon coupling, simulations of the Holstein model
also suffer from long autocorrelation times as a result of the
phonon-mediated, electron-electron binding. We introduce two additional
types of Monte Carlo updates, termed reflection and swap updates,
to address this issue. While similar types of updates have been employed
in DQMC simulations of the Holstein model, we are able to do so while
maintaining near linear scaling with system size.

Finally, we introduce techniques for efficiently measuring correlation functions. Elements of the matrix $M^{-1}(x)$ can be estimated stochastically, provide samples of the single-particle Green's function. It is frequently desirable to average correlation measurements over both real space and imaginary time to reduce the error. A straightforward approach to performing this average results in a computational cost that scales as $\mathcal{O}(N^2 L_\tau^2)$, which would violate our target of near linear-scaling cost. To recover the desired scaling, we formulated the real space and imaginary time averages as cross-correlations (with periodic boundaries), which enables their efficient evaluation using FFTs. As a consequence, measurements come almost ``for free,'' relative to the computational work required to sample the phonon field.

Electron-phonon interactions play an important role in describing emergent behaviors that occur in certain strongly interacting materials.
The methods outlined in this paper allow for the efficient simulation of
electron-phonon models over a much greater range of system sizes and parameter regimes, than was previously possible. This capability makes accessible the study of many new
material systems where electron-phonon interactions are believed to
play a prominent role in determining the low energy physics.

\begin{acknowledgments}
 B.~C.-S. was funded by a U.C. National Laboratory In-Residence Graduate Fellowship through the U.C. National Laboratory Fees Research Program. K.~B. acknowledges support from the center of Materials Theory as
a part of the Computational Materials Science (CMS) program, funded
by the U.S. Department of Energy, Office of Science, Basic Energy
Sciences, Materials Sciences and Engineering Division.
R.T.~S and O.~B acknowledge support from the U.S.~Department of Energy, Office of Science,
Office of Basic Energy Sciences, under Award Number DE-SC0022311.
C.~M acknowledges support from the U.S. Department of Energy, Office
of Science, Office of Advanced Scientific Computing Research, Department of Energy Computational
Science Graduate Fellowship under Award Number DE-SC0020347.
\end{acknowledgments}

\section*{Code availability}

The code is open source and available on Github \url{https://github.com/el-ph/el-ph}.

\appendix

\section{Review of path integral formalism\label{sec:path_integral}}

Here we review how the partition function for the Holstein model,
\begin{equation}
\mathcal{Z}=\trep e^{-\beta\hat{H}},
\end{equation}
can be formulated as a path integral over phonon fields. The trace
is over the combined Fock space for both electron and phonon operators.
The Suzuki-Trotter approximation yields~\cite{Trotter59}
\begin{align}
\mathcal{Z} & \approx\trep\left[e^{-\frac{\Delta_{\tau}}{2}\hat{H}_{\textrm{el-ph}}}e^{-\Delta_{\tau}\left(\hat{H}_{\mathrm{el}}+\hat{H}_{\mathrm{ph}}\right)}e^{-\frac{\Delta_{\tau}}{2}\hat{H}_{\textrm{el-ph}}}\right]^{L_{\tau}}\nonumber \\
 & =\trep\left[e^{-\Delta_{\tau}\hat{H}_{\textrm{el-ph}}}e^{-\Delta_{\tau}\hat{H}_{\mathrm{el}}}e^{-\Delta_{\tau}\hat{H}_{\mathrm{ph}}}\right]^{L_{\tau}},
\end{align}
where $\beta=\Delta_{\tau}L_{\tau}$ is the discretization in imaginary
time. This approximation is valid to order $\mathcal{O}(\Delta_{\tau}^{2})$. In the second step we used the fact that $\hat{H}_{\mathrm{ph}}$
and $\hat{H}_{\mathrm{el}}$ commute, and the cyclic property of the
trace.

\begin{widetext}

The next step is to evaluate the phonon trace in the position basis.
This is done by repeatedly inserting the identity operator $\int d^{N}x\,|x\rangle\langle x|,$
where $|x\rangle=|x_{1},x_{2},\dots x_{N}\rangle$ denotes an entire
real-space phonon configuration, such that the integral is understood
to be over all sites. Using $\langle x_{\tau}|x_{\tau+1}\rangle=\delta(x_{\tau}-x_{\tau+1})$,
the result is%

\begin{equation}
\mathcal{Z}\approx\tre\int\mathcal{D}x\,\prod_{\tau=0}^{L_{\tau}-1}e^{-\Delta_{\tau}\hat{H}_{\textrm{el-ph}}(x_{\tau})}e^{-\Delta_{\tau}\hat{H}_{\mathrm{el}}}\langle x_{\tau}|e^{-\Delta_{\tau}\hat{H}_{\mathrm{ph}}}|x_{\tau+1}\rangle,
\end{equation}
where the differential $\mathcal{D}x$ indicates a path integral over
all phonon fields $x_{i,\tau}$. $\hat{H}_{\textrm{el-ph}}(x_{\tau})$
denotes the operator $\hat{H}_{\textrm{el-ph}}$ with the replacement
$\hat{X}\mapsto x_{\tau}$, subject to the periodic boundary condition
$x_{L_{\tau}}\equiv x_{0}$. Next we write

\begin{equation}
\mathcal{Z}\approx\tre\int\mathcal{D}x\,e^{-S_{B}}\prod_{\tau=0}^{L_{\tau}-1}e^{-\Delta_{\tau}\hat{H}_{\textrm{el-ph}}(x_{\tau})}e^{-\Delta_{\tau}\hat{H}_{\mathrm{el}}},\label{eq:Z_2}
\end{equation}
where
\begin{equation}
e^{-S_{B}}=\prod_{\tau=0}^{L_{\tau}-1}\langle x_{\tau}|e^{-\Delta_{\tau}\hat{H}_{\mathrm{ph}}}|x_{\tau+1}\rangle.\label{eq:S_B_def}
\end{equation}
Again using a symmetric operator splitting,
\begin{equation}
e^{-\Delta_{\tau}\hat{H}_{\mathrm{ph}}}\approx e^{-\Delta_{\tau}\frac{\omega_{0}^{2}}{4}\hat{X}^{2}}e^{-\Delta_{\tau}\frac{1}{2}\hat{P}^{2}}e^{-\Delta_{\tau}\frac{\omega_{0}^{2}}{4}\hat{X}^{2}},
\end{equation}
we find
\begin{equation}
\langle x_{\tau}|e^{-\Delta_{\tau}\hat{H}_{\mathrm{ph}}}|x_{\tau+1}\rangle\approx e^{-\frac{\Delta_{\tau}\omega_{0}^{2}}{4}\left(x_{\tau}^{2}+x_{\tau+1}^{2}\right)}\langle x_{\tau}|e^{-\Delta_{\tau}\frac{1}{2}\hat{P}^{2}}|x_{\tau+1}\rangle\label{eq:brack_H_phi}
\end{equation}
which is locally valid to $\mathcal{O}(\Delta_{\tau}^{3})$. In this
notation, we are treating $x_{\tau}$ and $\hat{P}$ as $N$-component
vectors. The second factor can be evaluated by inserting a complete
set of momentum states,
\begin{align}
\langle x_{\tau}|e^{-\Delta_{\tau}\frac{1}{2}\hat{P}^{2}}|x_{\tau+1}\rangle & =\int d^{N}p\,\langle x_{\tau}|p\rangle e^{-\Delta_{\tau}\frac{1}{2}p^{2}}\langle p|x_{\tau+1}\rangle\nonumber \\
 & =\int d^{N}p\, e^{-\frac{\Delta_{\tau}}{2}p^{2}+ip\cdot(x_{\tau+1}-x_{\tau})}\nonumber \\
 & \propto e^{-\frac{\Delta_{\tau}}{2}\left(\frac{x_{\tau+1}-x_{\tau}}{\Delta_{\tau}}\right)^{2}}.\label{eq:bracket_P2}
\end{align}
Combining Eqs.~(\ref{eq:Z_2})--(\ref{eq:bracket_P2}), and recalling
that $x_{L_{\tau}}=x_{0}$, we arrive at the ``bosonic action''
for the phonons,
\begin{equation}
S_{B}\approx\Delta_{\tau}\sum_{i=1}^{N}\sum_{\tau=0}^{L_{\tau}-1}\left[\frac{1}{2}\omega_{0}^{2}x_{i,\tau}^{2}+\frac{\left(x_{i,\tau+1}-x_{i,\tau}\right)^{2}}{2\Delta{}_{\tau}^{2}}\right]+\mathrm{const}.
\end{equation}
This approximation is valid to order $\mathcal{O}(\Delta_{\tau}^{2})$
because we have chained the approximation in Eq.~(\ref{eq:brack_H_phi})
order $1/\Delta_{\tau}$ times.

With some algebraic rearrangement, the partition function in Eq.~(\ref{eq:Z_2})
may be written

\[
\mathcal{Z}\approx\int\mathcal{D}x\,e^{-\left(S_{{\rm B}}-\Delta_{\tau}\alpha\sum_{i,\tau}x_{i,\tau}\right)}\tre\prod_{\tau=0}^{L_{\tau}-1}\prod_{\sigma=\uparrow,\downarrow}e^{-\Delta_{\tau}\hat{V}_{\tau,\sigma}}e^{-\Delta_{\tau}\hat{K}_{\sigma}},
\]
\end{widetext}where
\begin{align}
\hat{V}_{\tau,\sigma} & =\sum_{i}\left(\alpha x_{i,\tau}-\mu\right)\hat{n}_{i,\sigma}\label{eq:Vhat}\\
\hat{K}_{\sigma} & =-\sum_{ij} t_{ij}\hat{c}_{i,\sigma}^{\dagger}\hat{c}_{j,\sigma},\label{eq:Khat}
\end{align}
are purely quadratic in the Fermions, making it possible to evaluate
the remaining electron trace. Since the two spin sectors are not coupled,
the result is~\cite{Blankenbecler81} 
\begin{align*}
\tre\prod_{\tau=0}^{L_{\tau}-1}\prod_{\sigma=\uparrow,\downarrow}e^{-\Delta_{\tau}\hat{V}_{\tau,\sigma}}e^{-\Delta_{\tau}\hat{K}_{\sigma}} & =\left(\det M\right)^{2}.
\end{align*}
where $M$ is a $NL_{\tau}\times NL_{\tau}$ matrix, conveniently
expressed in block form,

\begin{equation}
M\left(x\right)=\left(\begin{array}{ccccc}
I &  &  &  & B_{0}\\
-B_{1} & I\\
 & -B_{2} & \ddots\\
 &  & \ddots & \ddots\\
 &  &  & -B_{L_{\tau}-1} & I
\end{array}\right).
\end{equation}
$I$ is the $N\times N$ identity matrix, and 
\[
B_{\tau}=e^{-\Delta\tau V_{\tau}}e^{-\Delta_{\tau}K}.
\]
The $V_{\tau}$ and $K$ are matrix counterparts of the Fock-space operators
of Eqs.~(\ref{eq:Vhat}) and~(\ref{eq:Khat}), with elements
\[
\left(V_{\tau}\right)_{ij}=\delta_{ij}\left(\alpha x_{i,\tau}-\mu\right),\qquad K_{ij}=-t_{ij}.
\]

Putting together the pieces, the partition function may be approximated,
\begin{equation}
\mathcal{Z}\approx\int\mathcal{D}x\,e^{-\left(S_{{\rm B}}-\Delta_{\tau}\alpha\sum_{i,\tau}x_{i,\tau}\right)}\left(\det M\right)^{2},
\end{equation}
which is valid up to an error of order $\mathcal{O}\left(\Delta_{\tau}^{2}\right)$.

\section{Statistical symmetry of the action\label{sec:S_sym}}

Here we demonstrate how the particle-hole symmetry of the single-site
Holstein model at half-filling emerges in the action $S(x,\Phi_{\sigma})$
of Eq.~(\ref{eq:S_decompose}), provided that imaginary-time fluctuations
can be ignored.

Consider the change in action
\begin{equation}
\Delta S(x)=S(x)-S(x_{0}),
\end{equation}
for a move $x_{0}\rightarrow x$. For
particle-hole symmetry to be respected, we should find
\begin{equation}
\Delta S(x)\overset{?}{=}\Delta S(-x),\label{eq:DeltaS_sym_maybe}
\end{equation}
such that MC proposals $x_{0}\rightarrow x$ and $x_{0}\rightarrow-x$
would be accepted with equal probability. This condition is equivalent
to vanishing
\begin{equation}
\delta S=S(-x)-S(x).
\end{equation}
Observe that the starting configuration $x_{0}$ is irrelevant.
Let us now investigate the condition $\delta S=0$.

The bosonic action $S_{\mathrm{B}}(x)$ defined in Eq.~(\ref{eq:S_B})
is symmetric at half filling, but symmetry \emph{breaking} may arise
from the fermonic action $S_{\mathrm{F}}(x,\Phi_{\sigma})$ defined
in Eq.~(\ref{eq:S_F}). The result is,
\begin{equation}
\delta S=\frac{1}{2}\sum_{\sigma}\Phi_{\sigma}^{T}\left(D_{-x}^{-1}-D_{x}^{-1}\right)\Phi_{\sigma},
\end{equation}
where
\begin{equation}
D_{x}=A_{x}^{T}A_{x},
\end{equation}
and the auxiliary field $\Phi_{\sigma}$ is arbitrary. If $D_{x}=D_{-x}$, then $\delta S=0$, and the particle-hole symmetry
of Eq.~(\ref{eq:DeltaS_sym_maybe}) would be satisfied.

We now show that $D_{x}$ indeed satisfies this symmetry in the special
case of the adiabatic limit of the single-site Holstein model at half-filling
($\mu=0$). Without the hopping matrix $K$, the block matrices $B_{\tau}=e^{-\Delta_{\tau}\alpha x_{\tau}}$
become effectively scalar. In the absence of imaginary-time fluctuations,
we replace $B_{\tau}\rightarrow\bar{B}=e^{-\Delta_{\tau}\alpha\bar{x}}$.
Next, we explicitly calculate $A^{T}=\Lambda^{T}M^{T}$ using Eqs.~(\ref{eq:M_def})
and~(\ref{eq:Lambda_def}),
\begin{equation}
A_{\bar{x}}^{T}=\left(\begin{array}{ccccc}
\bar{B}^{1/2} &  &  &  & \bar{B}^{-1/2}\\
-\bar{B}^{-1/2} & \bar{B}^{1/2}\\
 & -\bar{B}^{-1/2} & \ddots\\
 &  & \ddots & \ddots\\
 &  &  & -\bar{B}^{-1/2} & \bar{B}^{1/2}
\end{array}\right).\label{eq:A_x}
\end{equation}
The subscript $\bar{x}$ emphasizes our neglect of imaginary-time
fluctuations. It follows,
\begin{equation}
D_{\bar{x}}=\left(\begin{array}{ccccc}
\bar{B}+\bar{B}^{-1} & -I &  &  & I\\
-I & \bar{B}+\bar{B}^{-1} & \ddots\\
 & -I & \ddots\\
 &  & \ddots & \ddots & -I\\
I &  &  & -I & \bar{B}+\bar{B}^{-1}
\end{array}\right),
\end{equation}
The transformation $\bar{x}\rightarrow-\bar{x}$ corresponds to $\bar{B}\rightarrow\bar{B}^{-1}$.
We conclude $D_{\bar{x}}=D_{-\bar{x}}$, as claimed, which implies
particle-hole symmetry of the action, Eq.~(\ref{eq:DeltaS_sym_maybe}).
The result is exact in the adiabatic limit (infinite atomic mass),
for which imaginary-time fluctuations can be ignored.

\section{Preconditioner implementation\label{sec:precond_impl}}

In Sec.~\ref{sec:Preconditioning} we described a preconditioner $P$
that is block diagonal in the Fourier space representation.
Along the diagonal, its $N\times N$ blocks have the form
\begin{equation}
\tilde{M}_{\omega,\omega}=I-e^{-i\phi_{\omega}}\bar{B},\label{eq:M_tilde_again}
\end{equation}
where
\begin{equation}
\phi_{\omega}=\frac{2\pi}{L_{\tau}}\left(\omega+\frac{1}{2}\right),\quad\bar{B}=e^{-\Delta_{\tau}\bar{V}}e^{-\Delta_{\tau}K} \label{eq:phi_scaling},
\end{equation}
and both $\bar{V}$ and $K$ are Hermitian matrices. Applying $P^{-1}$
to a vector requires application of the $N\times N$ matrices
$\tilde{M}_{\omega,\omega}^{-1}$, for all indices $\omega=0,1,\dots L_{\tau}-1$.
Here we describe how the kernel polynomial method (KPM)~\cite{Weisse06}
may be used to perform these matrix-vector products efficiently. This
approach systematically approximates each matrix $\tilde{M}_{\omega,\omega}^{-1}$
in polynomials of $\bar{B}$.

A first observation is that the matrices $e^{-\Delta_{\tau}\bar{V}}$
and $e^{-\Delta_{\tau}K}$ in their exact forms are positive definite
and Hermitian. From this, we can guarantee that all eigenvalues of
$\bar{B}$ are real~\cite{Drazin62}. The checkerboard approximation
to $e^{-\Delta_{\tau}K}$ slightly violates Hermiticity, but even
in this case, we have observed empirically that the eigenvalues of $\bar{B}$
remain exactly real in the context of our QMC simulations.

A second observation is that the eigenvalues $\bar{b}$ of $\bar{B}$
are bounded near $1$,
\begin{equation}
\bar{b}_{\mathrm{{\rm min}}}\leq\bar{b}\leq\bar{b}_{\max},
\end{equation}
otherwise $\Delta_{\tau}$ would not be sufficiently small
for the Suzuki-Trotter expansion to be meaningful. In the Holstein
model, $K$ will typically have a much larger spectral magnitude than
$\bar{V}$, so we can get the correct scaling with the approximation
$\bar{B}\approx e^{-\Delta_{\tau}K}$. On the square lattice with
hopping $t=1$, the extreme eigenvalues of $K$ are $\pm4$. Given
our choice of $\Delta_{\tau}=0.1$, the extreme eigenvalues will be
of order $\exp(\pm\Delta_{\tau}4)$, namely, $\bar{b}_{\mathrm{{\rm min}}}\approx0.7$
and $\bar{b}_{\max}\approx1.6$.

It will be convenient to define a rescaled matrix,
\begin{equation}
A=2(\bar{B}-\bar{b}_{\min})/\Delta\bar{b}-1,\label{eq:A_def}
\end{equation}
with $\Delta\bar{b}=\bar{b}_{{\rm max}}-\bar{b}_{\mathrm{{\rm min}}}$.
The eigenvalues $y$ of $A$ satisfy $-1\leq y\leq1$. This will allow
us to approximate
\begin{equation}
\tilde{M}_{\omega,\omega}^{-1}=\left(1-e^{-i\phi_{\omega}}\bar{B}\right)^{-1}=f_{\omega}(A),
\end{equation}
using Chebyshev polynomials in $A$. We may view
\begin{align}
f_{\omega}(y) & =\left(1-e^{-i\phi_{\omega}}\bar{b}\right)^{-1},\label{eq:f_y}
\end{align}
as a scalar function that acts on the eigenvalues $y$ of $A$, which
are related to the eigenvalues $\bar{b}$ of $\bar{B}$ via
\begin{equation}
y=2(\bar{b}-\bar{b}_{\min})/\Delta\bar{b}-1.\label{eq:y_to_b}
\end{equation}

\subsection{Chebyshev polynomial approximation}

An arbitrary scalar function $f(y)$ may be expanded in the basis
of Chebyshev polynomials,
\begin{equation}
f(y)=\sum_{m=0}^{\infty}c_{m}T_{m}(y),\label{eq:cheby}
\end{equation}
valid for $-1\leq y\leq1$. In this domain, the Chebyshev polynomials
can be written $T_{m}(y)=\cos\left(m\arccos y\right)$, such that
the coefficients $c_{m}$ may be interpreted as the cosine transform
of $f$ in the variable $\theta=\arccos(y)$.

The Chebyshev polynomials satisfy a generalized orthogonality relation,
\begin{equation}
\int_{-1}^{+1}w(y)T_{m}(y)T_{m'}(y)\mathrm{d}y=q_{m}\delta_{m,m'},
\end{equation}
where
\begin{align*}
w(y) & =\left(1-y^{2}\right)^{-1/2}\\
q_{m} & =\frac{\text{\ensuremath{\pi}}}{2}\left(1+\delta_{m,0}\right).
\end{align*}
The expansion coefficients are then given by
\begin{equation}
c_{m}=\frac{1}{q_{m}}\int_{-1}^{+1}w(y)T_{m}(y)f(y)\mathrm{d}y.\label{eq:c_integral}
\end{equation}
Usually a closed form solution for $c_{m}$ is not available, but
one can use Chebyshev-Gauss quadrature to obtain a good approximation
\begin{equation}
c_{m}\approx\frac{\pi}{q_{m} N_{\mathrm{Q}}}\sum_{n=0}^{N_{\mathrm{Q}}-1}\cos(m\theta_{n})f(\cos\theta_{n}),\label{eq:Cheby-Gauss}
\end{equation}
where $N_{\mathrm{Q}}$ is the number of quadrature points, and $\theta_{n}=\pi\left(n+\frac{1}{2}\right)/N_{\mathrm{Q}}$
are the abscissas. A fast Fourier transform can be used to calculate all coefficients $c_m$ efficiently~\cite{Weisse06}.

The utility of the expansion in Eq.~(\ref{eq:cheby}) is that we
can obtain a good approximation by truncating
\begin{equation}
f(y)\approx\sum_{m=0}^{N_{\mathrm{P}}-1}g_{m}c_{m}T_{m}(y),
\end{equation}
at an appropriate polynomial order $N_{\mathrm{P}}$. Here one has
the option to introduce damping factors $g_{m}$ associated with a
\emph{kernel}. The damping factors should be close to 1 for $m\ll N_{\mathrm{P}}$
and may decay to 0 as $m\rightarrow N_{\mathrm{P}}$. An appropriately
selected kernel guarantees uniform convergence of the Chebyshev series,
avoiding numerical artifacts such as Gibbs oscillations. In our application,
we are working with the smooth functions in Eq.~(\ref{eq:f_y}),
and we will simply set $g_{m}=1$.

For a given polynomial order $N_\mathrm{P}$, we find it sufficient to use $N_\mathrm{Q}=2N_\mathrm{P}$ quadrature points to approximate the expansion coefficients $c_{m}$ in Eq.~(\ref{eq:Cheby-Gauss}).

\subsection{Selecting the polynomial order}

\begin{figure}
\includegraphics[width=1.0\columnwidth]{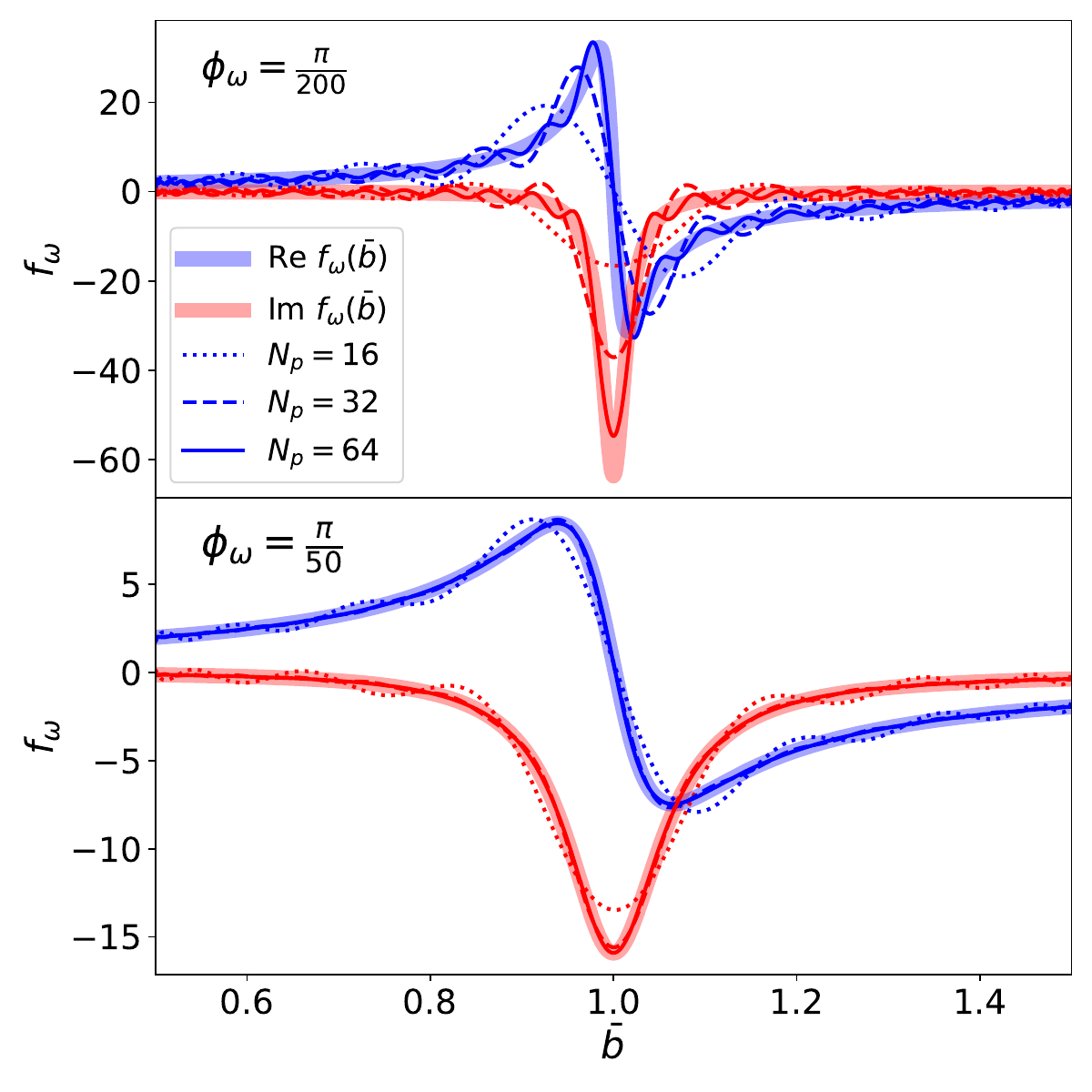}\caption{\label{fig:chebyshev}Chebyshev polynomial approximation of $f_{\omega}=(1-e^{-i\phi_{\omega}}\bar{b})^{-1}$
on a given interval $\bar{b}_{\mathrm{{\rm min}}}\protect\leq\bar{b}\protect\leq\bar{b}_{\max}$.
To resolve the sharp features in $f_{\omega}$ for small angles $\phi_{\omega}$,
the polynomial order should scale like $N_{\mathrm{P}}\sim\phi_{\omega}^{-1}$.}
\end{figure}

Figure~\ref{fig:chebyshev} illustrates Chebyshev approximation of
the real and imaginary parts of $f_{\omega}=(1-e^{-i\phi_{\omega}}\bar{b})^{-1}$
for various polynomial orders $N_{\mathrm{P}}$. Angles $\phi_{\omega}$
near zero give rise to sharper features in $f_{\omega}$, which require
a larger polynomial order $N_{\mathrm{P}}$ to resolve.

We will use the convention that the angle $\phi_\omega$
is between 0 and $\pi$. This effectively restricts our attention
to $0\leq\omega<L_{\tau}/2$, which is possible due to the symmetry
$f_{L_{\tau}-\omega-1}(\bar{b})=f_{\omega}^{\ast}(\bar{b})$.

In practice, we can achieve a good polynomial approximation using
the heuristic
\begin{equation}
N_{\mathrm{P}}=\left\lfloor \Delta\bar{b}\left(a_{1}\phi_{\omega}^{-1}+a_{2}\right)\right\rfloor ,\label{eq:NP_heuristic}
\end{equation}
where $\left\lfloor \cdot \right\rfloor$ denotes the floor function; the coefficients $a_{1}$ and $a_{2}$ are both of order $1$ and
independent of system details (temperature, etc.). Note that the polynomial
order $N_{\mathrm{P}}$ scales linearly with the range $\Delta\bar{b}=\bar{b}_{\mathrm{max}}-\bar{b}_{\mathrm{min}}$
over which an approximation is required. Observe that the polynomial
order $N_{\mathrm{P}}$ decays rapidly when $\omega$ moves away from
zero, such that the \emph{typical} value of $N_{\mathrm{P}}$ is of
order 1.

\subsection{Using KPM to evaluate matrix-vector products}

We wish to apply the matrix 

\begin{equation}
\tilde{M}_{\omega,\omega}^{-1}=(I-e^{-i\phi_{\omega}}\bar{B})^{-1}=f_{\omega}(A),
\end{equation}
to a vector, where $A$ is a rescaling of $\bar{B}$ as defined in
Eq.~(\ref{eq:A_def}). Using the truncated Chebyshev expansion, we
may approximate
\begin{equation}
\tilde{M}_{\omega,\omega}^{-1}\approx\sum_{m=0}^{N_{\mathrm{P}}-1}c_{m}T_{m}(A),\label{eq:f_A_approx}
\end{equation}
The expansion order $N_{p}$ and scalar coefficients $c_{m}$, given
in Eq.~(\ref{eq:c_integral}), implicitly depend on $\phi_{\omega}$,
$\bar{b}_{\mathrm{{\rm min}}}$, and $\bar{b}_{\mathrm{{\rm max}}}$.

A key result from KPM is that the task of evaluating the matrix-vector
product,
\begin{equation}
\tilde{M}_{\omega,\omega}^{-1}u\approx\sum_{m=0}^{N_{\mathrm{P}}-1}c_{m}T_{m}(A)u=\sum_{m=0}^{N_{\mathrm{P}}-1}c_{m}\alpha_{m},\label{eq:Minv_KPM}
\end{equation}
does \emph{not} require explicit construction of the dense matrix
$\tilde{M}_{\omega,\omega}^{-1}$. Instead, we will iteratively calculate
the vectors
\begin{equation}
\alpha_{m}=T_{m}(A)u.
\end{equation}
The Chebyshev polynomials satisfy a two-term recurrence relation,
\begin{equation}
T_{m+1}(A)=2AT_{m}(A)-T_{m-1}(A).\label{eq:cheby_recur}
\end{equation}
Multiplying by $u$ on the right yields an explicit scheme for computing
$\alpha_{m}$,
\begin{equation}
\alpha_{m+1}=2A\alpha_{m}-\alpha_{m-1},\label{eq:alpha_recur}
\end{equation}
beginning with
\begin{equation}
\alpha_{0}=u,\quad\alpha_{1}=Au.
\end{equation}
As the vectors $\alpha_{m}$ become available, they are accumulated
into the right-hand side of Eq.~(\ref{eq:Minv_KPM}), eventually
giving the desired matrix-vector product.

\subsection{A full recipe for the preconditioner}

Here we summarize all steps needed to apply the preconditioner
in Eq.~(\ref{eq:P_def}) efficiently. Our task is to evaluate the matrix-vector
product,

\begin{align}
P^{-1}v & =\mathcal{U}^{\dagger}\tilde{P}^{-1}\mathcal{U}v.
\end{align}
The unitary matrix $\mathcal{U}$ is defined in Eq.~(\ref{eq:udef})
and can be efficiently applied with an FFT. The matrix $\tilde{P}$
is zero except for its diagonal blocks $\tilde{M}_{\omega,\omega}$, which are given by Eq.~(\ref{eq:M_tilde_again}).
The main challenge is to apply the $N\times N$ matrix $\tilde{M}_{\omega,\omega}^{-1}$
to a vector. We must do so for each index $\omega$.

The matrix $\tilde{M}_{\omega,\omega}$ is a function of $\bar{B}=e^{-\Delta_{\tau}\bar{V}}e^{-\Delta_{\tau}K}$. If we can find numbers $\bar{b}_{\mathrm{min}}$
and $\bar{b}_{\mathrm{max}}$ that \emph{assuredly} bound all eigenvalues
of $\bar{B}$, then we may approximate $\tilde{M}_{\omega,\omega}^{-1}$ as in Eq.~(\ref{eq:f_A_approx}).

To estimate $\bar{b}_{\mathrm{max}}$, we may use the Arnoldi iteration,
repeatedly applying the matrix $\bar{B}$ to an initial random vector.
This method produces an upper Hessenberg matrix, which serves as a
low-rank approximation to $\bar{B}$. After about 20 iterations, the
largest eigenvalue of this Hessenberg matrix (increased by 5\%, to
be safe) provides a suitable estimate of $\bar{b}_{\mathrm{max}}$.
For numerical stability reasons, we estimate $\bar{b}_{\mathrm{min}}$
by applying the Arnoldi iteration to $\bar{B}^{-1}=e^{\Delta_{\tau}K}e^{\Delta_{\tau}\bar{V}}$,
estimating its maximum eigenvalue and then taking the inverse. This
is possible because, just like for $\bar{B}$, we are able to
apply $\bar{B}^{-1}$ to a vector efficiently.

Given the approximation in Eq.~(\ref{eq:f_A_approx}), we can efficiently
calculate $\tilde{M}_{\omega,\omega}^{-1}u$ using Eq.~(\ref{eq:Minv_KPM}),
where the vectors $\alpha_{m}=T_{m}(A)u$ are iteratively calculated
using the Chebyshev recurrence in Eq.~(\ref{eq:alpha_recur}).

The appropriate polynomial order $N_{\mathrm{P}}$ depends on the
index $\omega$. A reasonable choice is given in Eq.~(\ref{eq:NP_heuristic}).

\subsection{Scaling of costs\label{subsec:precond_scaling}}

The calculation of the matrix-vector product in Eq.~(\ref{eq:Minv_KPM})
requires $N_{\mathrm{P}}-1$ matrix-vector multiplications involving
$\bar{B}$, where $N_{\mathrm{P}}$ depends
on $\omega$ via Eq.~(\ref{eq:NP_heuristic}). Since the indices
$\omega$ and $L-\omega-1$ are effectively equivalent, we restrict
attention to $0\leq\omega<L_{\tau}/2$. We can sum over all such $\omega$
values to count the total number of required matrix-vector multiplications
\begin{align}
N_{\textrm{mat-vec}} & =2\sum_{\omega=0}^{L_{\tau}/2-1}\left[N_{\mathrm{P}}(\omega)-1\right]\nonumber \\
 & =2\sum_{\omega=0}^{L_{\tau}/2-1}\left\lfloor \Delta\bar{b}\left(a_{1}\phi_{\omega}^{-1}+a_{2}\right)\right\rfloor -L_{\tau}.
\end{align}
The factor of 2 accounts for the skipped indices, $L_{\tau}/2\leq\omega<L_{\tau}$.
Removing the floor function is justified when $\omega$ is order 1, such that $\phi_{\omega}^{-1}$ is order $L_\tau$ (cf. Eq.~(\ref{eq:phi_scaling})), and in general produces an \emph{upper
}bound,
\begin{align}
N_{\textrm{mat-vec}} & \leq2\Delta\bar{b}\left(a_{1}\sum_{\omega=0}^{L_{\tau}/2-1}\phi_{\omega}^{-1}+a_{2}L_{\tau}/2\right)-L_{\tau}.
\end{align}
We can explicitly evaluate the sum,
\begin{equation}
\sum_{\omega=0}^{L_{\tau}/2-1}(\omega+1/2)^{-1}=\ln4+\gamma+\psi(L_{\tau}/2+1/2),
\end{equation}
where $\gamma=0.577\dots$ is the Euler-Mascheroni constant and $\psi(x)=\ln x+\mathcal{O}(x^{-1})$
is the digamma function. To a good approximation, the upper bound
is 
\begin{align}
N_{\textrm{mat-vec}} & \lesssim L_{\tau}\Delta\bar{b}\left[\frac{a_{1}}{\pi}\left(\gamma+\ln2L_{\tau}\right)+a_{2}\right]-L_{\tau}.\label{eq:N_mv_bound}
\end{align}

Typically $a_{1}=a_{2}=1$ and $\Delta\bar{b}\approx1$. For, say,
$L_{\tau}=200$ (corresponding to inverse temperature $\beta=20$
at $\Delta_{\tau}=0.1$), the bound of Eq.~(\ref{eq:N_mv_bound})
gives,
\begin{equation}
N_{\textrm{mat-vec}}/L_{\tau}\lesssim2.1,
\end{equation}
whereas direct numerical evaluation of the sum yields $N_{\textrm{mat-vec}}/L_{\tau}=1.6$.
We infer that the bound of Eq.~(\ref{eq:N_mv_bound}) is in general
a fairly tight one.

Note that $L_{\tau}$ applications of the matrix $\bar{B}=\exp(-\Delta_{\tau}\bar{V})\exp(-\Delta_{\tau}K)$
are equivalent to the work required to apply the matrix $M$ in Eq.~(\ref{eq:M_def}).
It follows that the task of applying the preconditioner in the Fourier
basis, $\tilde{P}^{-1}$, is about two times more expensive than applying
$M$. To apply $P^{-1}=\mathcal{U}^{\dagger}\tilde{P}^{-1}\mathcal{U}$, we additionally
require two FFTs. For the benchmarks performed in this paper, we measured
numerically that the total cost to apply $P^{-1}$ is about three times
greater than the cost to apply $M$.

\begin{figure}
\includegraphics[width=1.0\columnwidth]{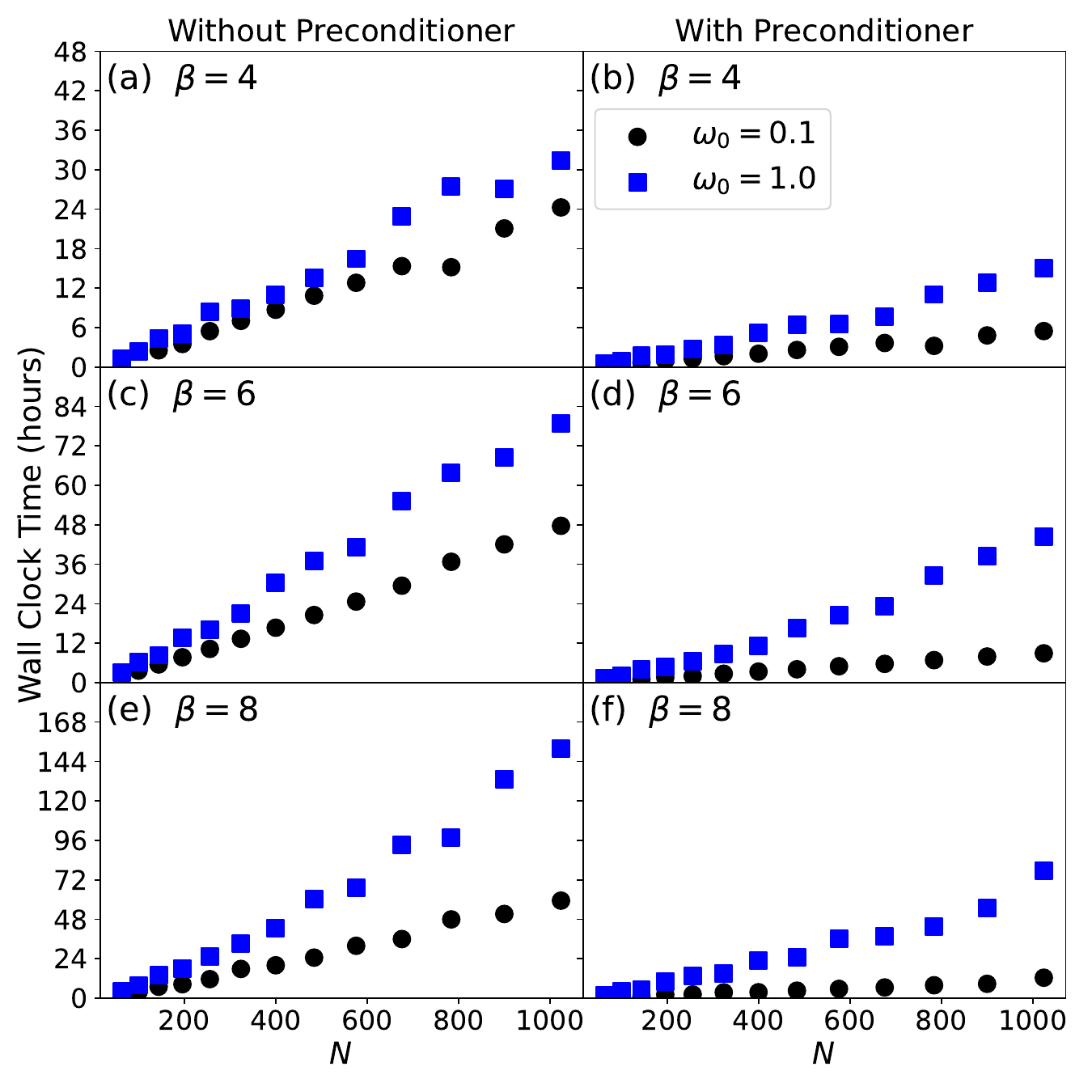}
\caption{\label{fig:time_vs_N}Wall clock time for a full simulation (including over 650k CG solves) as a function of
system size $N$. The corresponding average iteration counts per CG solve are shown in Fig.~\ref{fig:iters_vs_N}.}
\end{figure}

\begin{figure}
\includegraphics[width=1.0\columnwidth]{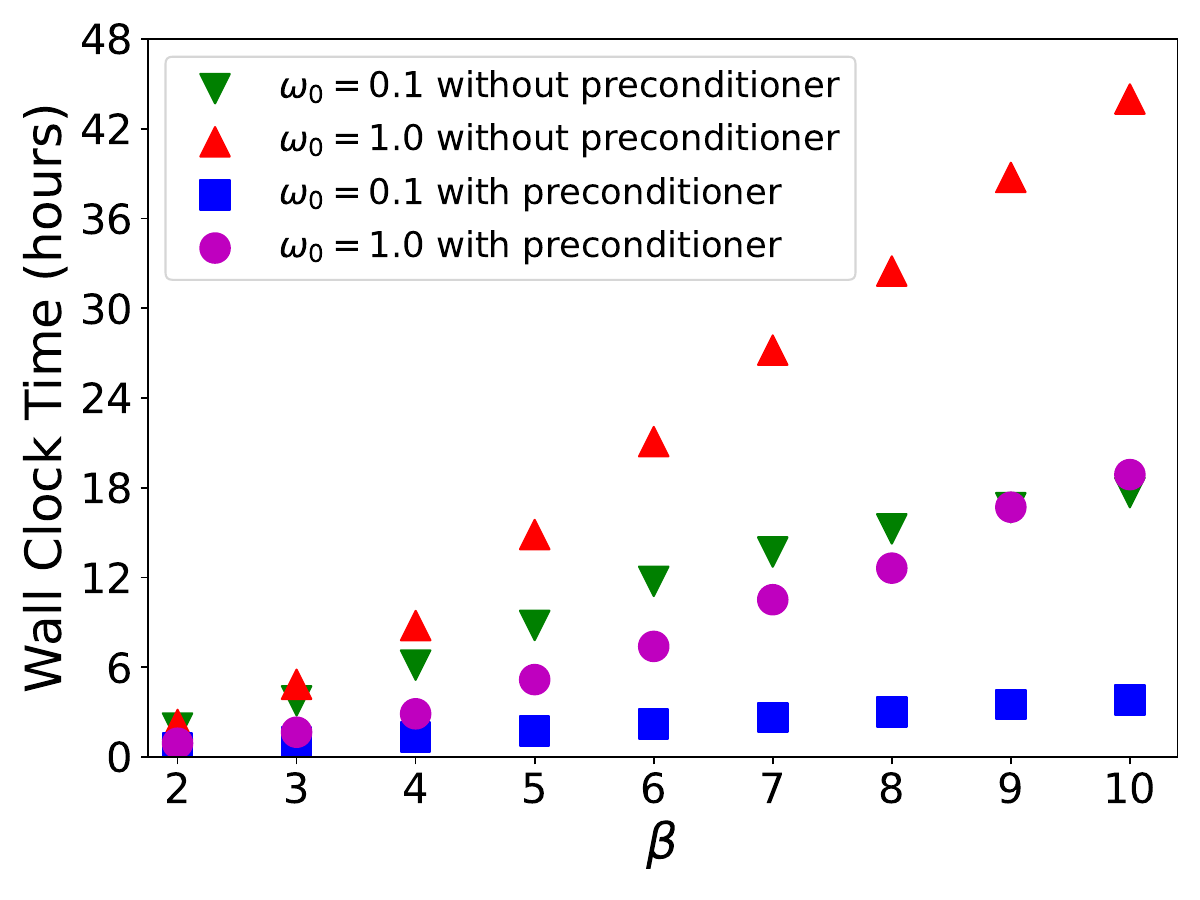}
\caption{\label{fig:time_vs_beta}Wall clock time for a full simulation as a function of
inverse temperature $\beta$ with system size $N=256$. The corresponding average iteration counts are shown in Fig.~\ref{fig:iters_vs_beta}.}
\end{figure}

\section{Simulation time versus system size and inverse temperature\label{sec:wall_clock_time}}

In this appendix we report the wall clock time for a full simulation, as a function of both system size $N$ and inverse temperature $\beta$. Each simulation was performed using only a single core of an Intel i7-4770 and i7-2600 processor (no parallelism).

All simulations used to generate results in this appendix were for Holstein systems with a dimensionless electron-phonon coupling of $\lambda=0.25$. Each simulation performed $N_{\rm therm}=1000$ HMC updates to equilibrate the system, followed by an additional $N_{\rm sim}=2000$ HMC updates.
Each HMC update consisted of $N_t=100$ time-steps, and each time-step requires two CG solves.
Each HMC update was followed by $4$ reflection and $4$ swap updates, requiring $8+8$ CG solves. Additionally, a total of $N_{\rm sim}=2000$ measurements were taken, each requiring $N_{\rm rv}=10$ CG solves. In total, the simulation involved approximately 668k CG solves, which comprise the dominant computational cost.
This simulation run-time was sufficient to achieve very accurate statistics, as demonstrated by the $S_\mathrm{CDW}$ measurements shown in Fig.~\ref{fig:iters_vs_beta}(a).

Figure~\ref{fig:time_vs_N} displays the total simulation wall clock time as a function of $N$, and corresponds to Fig.~\ref{fig:iters_vs_N}, which shows the average iteration count per CG solve. In all panels we see that the wall clock time scales in an approximately linear fashion with $N$. Empirical fitting of the wall clock time to a power law curve in $N$ yields an exponent between 1.0 and 1.3 in all cases. Additionally, we see that the preconditioner uniformly decreases the simulation time, although the relative speed-up is more significant at $\omega_0=0.1$ than $\omega_0=1.0$.

In similar fashion, Fig.~\ref{fig:time_vs_beta} shows the wall clock time versus $\beta$, and should be compared with Fig.~\ref{fig:iters_vs_beta}(b), which reports the average iteration count per CG solve. Once again we see that the preconditioner strictly reduces total simulation times, and that the wall clock time scales near linearly with $\beta$.

\bibliographystyle{unsrt}
\bibliography{HMC_Paper.bib}

\end{document}